\documentclass[fleqn,twoside]{article}
\usepackage{espcrc2}
\usepackage{axodraw}

\usepackage{graphicx}
\usepackage[figuresright]{rotating}

\newcommand{\AmS}{{\protect\the\textfont2
  A\kern-.1667em\lower.5ex\hbox{M}\kern-.125emS}}

\title{Recent Progress in Perturbative Quantum Field Theory\thanks{Plenary 
   talk at XXXI International Conference on High Energy Physics (ICHEP), 
      Amsterdam, July 2002.}}

\author{Zvi Bern\address[MCSD]{Department of Physics and Astronomy \\ 
        University of California at Los Angeles \\
        Los Angeles, CA 90095}%
        \thanks{This work was supported by the US Department of Energy 
                under grant DE-FG03-91ER40662.}}

\def\lsim{\buildrel < \over {_\sim}}

\def\sect#1{Section~\ref{#1}}
\def\fig#1{Figure~\ref{#1}}
\def\figs#1#2{Figure~\ref{#1}~and~\ref{#2}}
\def\Ord{{\cal O}}

\begin{document}

\begin{abstract}

In this talk, the recent breakthrough in two-loop perturbative
calculations is reviewed, with emphasis on the applications to
phenomenological studies.  The recent precision measurement of the
anomalous magnetic moment of the muon by the Brookhaven $g-2$
collaboration is also used as an illustrative example of the importance of
Standard Model precision perturbative calculations to further our
understanding of Nature.
\vspace{1pc}
\end{abstract}

\maketitle

\section{INTRODUCTION}

Precision calculations in the Standard Model have a long history
dating back to Schwinger's original QED calculations.  The importance
of precision Standard Model calculations in modern day particle
physics was recognized by the 1999 Nobel Prize awarded to
Gerard `t~Hooft and Martinus Veltman.  The tools and theoretical
foundation provided by `t~Hooft and Veltman have led to a plethora of
precision calculations, including the famous `blueband' plot, shown
many times at this conference, bounding the mass of a Standard Model
Higgs boson, $m_H \lsim 196$--$230$ GeV at 95\%
CL~\cite{HiggsRadCorr}.

In this talk we describe recent progress in precision calculations, as
well as present examples illustrating the importance of higher loop
computations for comparing theory to experiments.  The most
spectacular recent example of this is the precision
measurement of the anomalous magnetic moment of the muon by the
Brookhaven $g-2$ collaboration~\cite{GM2,GM22,Semertzidis,Teubner}.
This experiment is an ideal example due to its impressive
sensitivity to the quantum corrections arising from all three
components of the Standard Model: Quantum Electrodynamics, Electroweak
Theory, and Quantum Chromodynamics.  The incredible agreement of
approximately seven digits of accuracy between theory and experiment
is a testament to the remarkable experimental and theoretical efforts
that have gone into this attempt to discover new physics beyond the
Standard Model.

This talk also focuses on the very recent breakthrough in our ability
to calculate two-loop quantum corrections when more than a single
kinematic invariant is present.  This breakthrough is the culmination
of many years of theoretical effort.  To illustrate the promise of
this breakthrough, we present two very recent applications improving
our understanding of Higgs physics, which is of central importance in
particle physics. As one example, just two weeks before the
conference, Anastasiou and Melnikov~\cite{AnastasiouMelnikov} computed
the exact NNLO result for next-to-next-to-leading order (NNLO) QCD
corrections to inclusive Higgs production, in perfect agreement with
the earlier series expansion result of Harlander and
Kilgore~\cite{HarlanderKilgore}.  The second example, completed a
month prior to the conference, is gluon fusion into a photon
pair~\cite{BDDgggamgam,Hbkgd}.  This is a significant component of the
background to the preferred search mode for discovering and measuring
the Higgs mass at the Large Hadron Collider, if the Higgs is light
($M_H < 140$ GeV), as suggested by precision electroweak measurements.

These examples represent initial steps in applying the breakthrough to
phenomenology.  It is now clear that over the next few years, NNLO
calculations of jet observables at both hadron and lepton colliders
will become available~\cite{GOTY2to2,BDDgggg,NNLOee}. (Other related
issues may be found in Frixione's~\cite{FrixioneTalk} QCD theory
talk.)  Precision calculations of Bhabha scattering is another example
that will surely benefit from the
breakthrough~\cite{BhabhaTwoLoop,GloverBhabha}. This process is
important for measuring luminosity at $e^+ e^-$ colliders. More
generally, many observables which could not previously be calculated
to the required order of perturbation theory necessary to match
experimental precisions will now be computed.

This talk is organized as follows. In \sect{AnomalousMuonSection} the
theoretical calculations required for a precision prediction of the
anomalous magnetic moment of the muon are reviewed briefly.  In
\sect{IdeasSection} some examples illustrating the clever tricks that
enter challenging perturbative calculations are given.  A brief
overview of the status of loop calculations is presented in
\sect{OverviewSection}.  In \sect{BreakthroughSection} we discuss the recent
breakthrough in our ability to compute quantities of interest with 
more than a single kinematic variable.
In \sect{ExamplesSection} two
very recent examples of applications of the new advances to
phenomenology are presented.  Prospects for the future are described
in \sect{SummarySection}.

\section{EXAMPLE: ANOMALOUS MAGNETIC MOMENT CALCULATIONS}
\label{AnomalousMuonSection}

The anomalous magnetic moment of the muon is an especially interesting
quantity because of its sensitivity to high scale new physics. It is
about $40,000 \simeq (m_\mu/ m_e)^2$ times more sensitive than the
electron magnetic moment in most potential scenarios of physics beyond
the Standard Model. 
The incredible precision of both the theoretical
predictions and the experiments, coupled with its sensitivity to
potential new physics makes measurement of the anomalous magnetic
moment of the muon an important means of searching for new physics
beyond the Standard Model 
(see {\it e.g.} refs.~\cite{GM2Update,MarcianoHarbinger}).

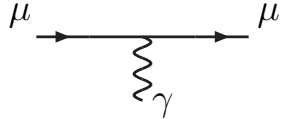
\begin{figure}
\SetScale{.8}  
\begin{center}
\begin{picture}(130,40)(0,0)
\SetWidth{1.3}
\ArrowLine(20,40)(45,40)
\Text(10,40)[c]{\footnotesize{\Large$\mu$}}
\Line(45,40)(95,40)
\ArrowLine(95,40)(120,40)
\Text(104,40)[c]{\footnotesize{\Large$\mu$}}
\Photon(70,10)(70,40){4}{3}
\Text(64,6)[c]{\footnotesize{\Large$\gamma$}}
\end{picture}
\end{center}
\vskip -1 cm 
\caption{The Feynman diagram describing the Dirac prediction for the
magnetic moment of the muon.}
\label{DiracFigure}
\end{figure}

\begin{figure}
\begin{center}
\SetScale{.8}  
\begin{picture}(130,50)(0,0)
\SetWidth{1.5}
\ArrowLine(20,40)(45,40)
\Text(10,40)[c]{\footnotesize{\Large$\mu$}}
\Line(45,40)(95,40)
\ArrowLine(95,40)(120,40)
\Text(104,40)[c]{\footnotesize{\Large$\mu$}}
\PhotonArc(70,40)(25,0,180){4}{6.5}
\Text(600,55)[c]{\footnotesize{\Large$\gamma$}}
\Photon(70,10)(70,40){4}{3}
\Text(64,6)[c]{\footnotesize{\Large$\gamma$}}
\end{picture}
\end{center}
\vskip -1 cm 
\caption{The Feynman diagram containing the 
leading quantum correction to the magnetic moment
of a lepton, first obtained by Schwinger.}
\label{SchwigerFigure}
\end{figure}
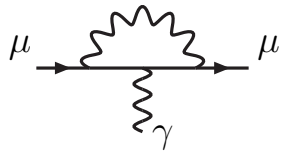

The magnetic moment of an elementary particle is related to it spin
via the gyromagnetic ratio:
\begin{equation}
\vec \mu = g \, {e \over 2 m } \vec S \,,
\end{equation}
where $\vec \mu$ is the magnetic moment and $ \vec S$ the spin.
From the Dirac equation, the gyromagnetic ratio is $g=2$ for
spin 1/2 particles such as the muon or the electron.  The tree-level Feynman
diagram describing the Dirac prediction is depicted in
\fig{DiracFigure}.  Quantum mechanical effects represented by loop
diagrams such as the one in \fig{SchwigerFigure}, however, alter this
result very slightly leading to a non-vanishing `anomalous' magnetic
moment:
\begin{equation}
a_\mu = {g-2 \over 2}\,.
\end{equation}
Computations of anomalous magnetic moments have a long history dating
back to Schwinger's original calculation of the $\Ord(\alpha)$
contribution in 1948~\cite{Schwinger}. 

The incredibly precise experimental result announced at this
conference by Yannis Semertzidis~\cite{Semertzidis} on behalf of the
Brookhaven muon $g-2$ collaboration~\cite{GM22} is
\begin{equation}
a_\mu^{\rm exp} = 11 \, 659\, 204 (7)(5) \times 10^{-10}\,,
\label{BNLResult}
\end{equation}
which is a significant improvement over the previous measurement~\cite{GM2},
\begin{equation}
{a_\mu^{\rm exp} = 11 \, 659 \, 202 (15) \times 10^{-10}}\,,
\end{equation}
and an even greater improvement on the series of experiments at CERN that
ended in 1977. 

\begin{figure}[htb]
\begin{center}
\SetScale{.8}  
\begin{picture}(110,57)(0,0)
\SetWidth{1.9}
\ArrowLine(20,40)(50,40)
\Text(15,41)[c]{\footnotesize{\Large $\mu$}}
\Line(45,40)(95,40)
\ArrowLine(95,40)(120,40)
\Text(100,41)[c]{\footnotesize{\Large $\mu$}}
\DashCArc(70,40)(25,0,180){3}
\Text(60,62)[c]{\footnotesize{\Large $\tilde \nu$}}
\Photon(70,10)(70,40){4}{3}
\Text(65,5)[c]{\footnotesize{\Large $\gamma$}}
\end{picture}
\end{center}
\vskip -1. cm 
\caption{New physics can lead to a contribution to the anomalous
magnetic moment beyond that found in the Standard Model. The dotted
line labeled by $\tilde \nu$ represents a particle of a New
Physics scenario.}
\label{NewPhysicsFigure}
\end{figure}
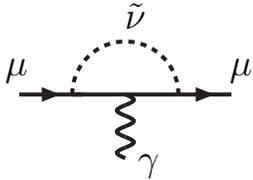

A discrepancy between the experimental result and the theoretical
prediction using the Standard Model can in principle indicate new
physics beyond the Standard Model,
\begin{equation}
{a_\mu^{\rm exp} - a_\mu^{\rm SM} = a_\mu^{\rm new\ physics}}\,.
\end{equation}
In \fig{NewPhysicsFigure} a sample diagram which could lead to a New
Physics contribution to the anomalous magnetic moment is shown. The
virtual $\tilde \nu$ particle depicted in the diagram
would be some as yet undiscovered particle coming from a New Physics 
scenario, such as supersymmetry.

The initial Brookhaven~\cite{GM2} result caused some excitement in the
community~\cite{MarcianoHarbinger} because of a {$2.6 \sigma$}
discrepancy between theory and experiment.  However, as described
below in the subsection on hadronic contributions, a correction was
made in the theoretical computation that significantly reduces the
discrepancy.  Amusingly, the improved experimental results announced
at this conference bring back the discrepancy to essentially the
original level.  However, the theoretical uncertainties associated
with the hadronic corrections are still the subject of intense debate
and it is not yet clear if the discrepancy is due to new physics.

For the purposes of this talk what is important is not the small
discrepancy between theory and experiment, but the astonishing
agreement to better than six significant digits. This agreement is a
great triumph for modern particle physics, from both the experimental
and theoretical sides.

On the theoretical side the computation requires a firm grasp of all
components of the Standard Model.  We now briefly review the intense
theoretical effort that has gone into producing Standard Model
predictions with the requisite precision. More thorough reviews may be
found elsewhere~\cite{GM2Update,MarcianoHadronic,Teubner}.

\subsection{QED contributions}

Calculations of the QED contributions to the anomalous magnetic moment
of the muon and other leptons have a long history (see
e.g. refs.~\cite{QEDMoment,Kinoshita4Loop,LaportaQED}) starting with
the seminal work of Schwinger~\cite{Schwinger}.  The QED effects are
by far the largest of the contributions, but are under the best
theoretical control.

\begin{figure}[htb]
\begin{picture}(100,50)(0,0)
\SetScale{.8}
\SetWidth{1.5}
\Line(20,40)(120,40)
\PhotonArc(60,40)(25,0,180){4}{6.5}
\PhotonArc(80,40)(25,0,180){4}{6.5}
\Photon(70,10)(70,40){4}{3}
%
%
\SetOffset(110,0)
\Line(20,40)(120,40)
\PhotonArc(70,40)(33,0,60){4}{3}
\PhotonArc(70,40)(33,120,180){4}{3}
\Oval(70,67)(10,18)(0)
\Photon(70,10)(70,40){4}{3}
%
\end{picture}
\vskip -1 cm 
\caption{Examples of two-loop QED contributions to the anomalous
magnetic moment.}
\end{figure}
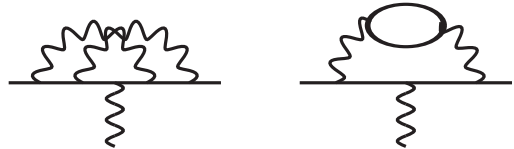

The combined results of many years of effort is that the QED
contribution is
\begin{eqnarray}
a_\mu^{\rm QED}\! & = & \! {{\alpha \over 2 \pi}}
+ 0.765857376(27) {\Bigl( {\alpha \over \pi}\Bigr)^2} \nonumber \\
&& + 24.05050898(44) {\Bigl( {\alpha \over \pi}\Bigr)^3}   \\
&&  + 126.07(41) {\Bigl( {\alpha \over \pi} \Bigr)^4} \nonumber \\
&& + 930(170)  {\Bigl( {\alpha \over \pi} \Bigr)^5}  
+ \cdots  \nonumber
\end{eqnarray}
These results are from the theory update of Czarnecki and
Marciano~\cite{GM2Update}, compiling the results of
refs.~\cite{QEDGM2,LaportaQED}. The four-loop $(\alpha/\pi)^4$
contributions, for example, consist of a total of 891 Feynman diagrams
which were evaluated by Kinoshita and his
collaborators~\cite{Kinoshita4Loop} via numerical means.  (A recent
correction actually shifts this very slightly, but with
negligible effect on the prediction of the muon anomalous magnetic
moment~\cite{KinoshitaCorrect}.)  The five loop $(\alpha/\pi)^5$
corrections are not exact calculations but are instead estimated using
various approximations~\cite{QEDFiveLoop}.  An important contribution
which significantly reduces the uncertainty is the exact analytical
result for the three-loop computation~\cite{LaportaQED}.  The net
result for the QED contribution to the anomalous magnetic moment of
the muon is:
\begin{equation}
a_\mu^{\rm QED} = 11\, 658\, 470.5 9 (29) \times 10^{-10}\,.
\end{equation}
Although the QED contribution is the largest of the contributions, the
theoretical uncertainty is tiny compared to the experimental
uncertainty in eq.~(\ref{BNLResult}) due to the high order of these
calculations.

\subsection{Electroweak contributions}

\begin{figure}[htb]
\begin{picture}(200,55)(0,0)
\SetScale{.8}  
\SetWidth{1.5}
\Line(20,40)(120,40)
\Text(10,30)[c]{\footnotesize{\Large$\mu$}}
\Text(104,30)[c]{\footnotesize{\Large$\mu$}}
\PhotonArc(60,40)(25,0,180){4}{6.5}
\Text(35,61)[c]{\footnotesize{\Large$\gamma$}}
\PhotonArc(80,40)(25,0,180){4}{6.5}
\Text(80,61)[c]{\footnotesize{\large$W$}}
\Photon(70,10)(70,40){4}{3}
\Text(64,7)[c]{\footnotesize{\Large$\gamma$}}
%
%
\SetOffset(100,10)
\SetWidth{1.5}
\ArrowLine(20,70)(45,70)
\Text(9,56)[c]{\footnotesize{\Large$\mu$}}
\Line(45,70)(95,70)
\ArrowLine(95,70)(120,70)
\Text(105,56)[c]{\footnotesize{\Large$\mu$}}
\Text(65,65)[c]{\footnotesize{\Large$\nu_\mu$}}
\Photon(45,70)(45,40){3}{3}
\Photon(95,70)(95,40){3}{3}
\ArrowLine(45,40)(95,40)
\ArrowLine(95,40)(70,15)
\ArrowLine(70,15)(45,40)
\Text(86,44)[c]{\footnotesize{\Large$\gamma$}}
\Text(26,44)[c]{\footnotesize{\Large$Z$}}
\Photon(70,-10)(70,15){3}{2}
\Text(64,-5)[c]{\footnotesize{\Large$\gamma$}}
%
\end{picture}
\vskip -1. cm 
\caption{Two of the 1650 two-loop electroweak Feynman diagrams 
contributing to the anomalous magnetic moment of the muon.}
\label{TwoloopElectroweakFigure}
\end{figure}
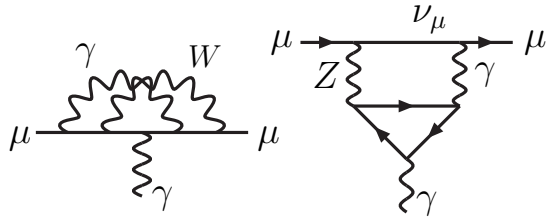

The electroweak contribution is much smaller than the QED contribution
because of the suppression due to $W$ or $Z$ propagators.
Nevertheless, the Brookhaven experiment is sensitive to these tiny
corrections.  The one-loop electroweak corrections displayed in
\fig{OneloopElectroweakFigure} were computed about 30 years
ago~\cite{Electroweak1Loop}.  The more recently computed two-loop
corrections~\cite{Electroweak2Loop} are surprisingly large, shifting
the one-loop results by about 23\%.  This large shift is due mainly to
the appearance of large logarithms of the form $\ln(M_Z^2/M_\mu^2)$.
The two-loop electroweak contribution is non-trivial to calculate since
it is described by a total of 1650 Feynman diagram of which about 200
are important.  Two sample diagrams are depicted in
\fig{TwoloopElectroweakFigure}. The final result for the electroweak
contributions including the two loop corrections
is~\cite{Electroweak2Loop}
\begin{equation}
a_\mu^{\rm EW} = 15.1(4)\times 10^{-10}.
\end{equation}
The theoretical uncertainty in this quantity is tiny compared to the
uncertainty in the experimental measurement (\ref{BNLResult}).
However, the overall contribution of the expression is well within the
sensitivity of the experiment, providing an exquisitely indirect
confirmation of electroweak theory~\cite{GM22,Semertzidis}.

\begin{figure}[htb]
\begin{center}
\begin{picture}(200,60)(0,0)
\SetScale{.7}  
\SetWidth{1.5}
\ArrowLine(20,40)(45,40)
\Text(9,34)[c]{\footnotesize{\Large$\mu$}}
\Line(45,40)(95,40)
\ArrowLine(95,40)(120,40)
\Text(92,34)[c]{\footnotesize{\Large$\mu$}}
\PhotonArc(70,40)(25,0,180){4}{6.5}
\Text(57,58)[c]{\footnotesize{\large$Z$}}
\Photon(70,10)(70,40){4}{3}
\Text(60,6)[c]{\footnotesize{\Large$\gamma$}}
%
%
\SetOffset(100,0)
\SetWidth{1.5}
\ArrowLine(20,70)(45,70)
\Text(10,57)[c]{\footnotesize{\Large$\mu$}}
\Line(45,70)(95,70)
\ArrowLine(95,70)(120,70)
\Text(90,57)[c]{\footnotesize{\Large$\mu$}}
\Text(50,60)[c]{\footnotesize{\Large$\nu_\mu$}}
\PhotonArc(70,70)(25,180,0){4}{6.5}
\Text(28,30)[c]{\footnotesize{\large$W$}}
\Text(73,30)[c]{\footnotesize\large{$W$}}
\Photon(70,10)(70,40){4}{3}
\Text(59,6)[c]{\footnotesize{\Large$\gamma$}}
%
%
\end{picture}
\end{center}
\vskip -1 cm 
\caption{One-loop electroweak Feynman diagrams contributing to the anomalous
magnetic moment of the muon.}
\label{OneloopElectroweakFigure}
\end{figure}
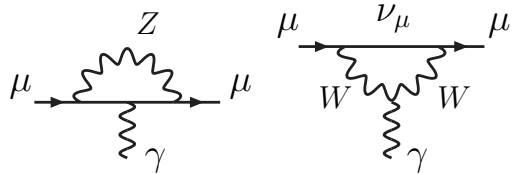

\subsection{Hadronic contributions}

The most problematic of the contributions are the hadronic corrections
to the anomalous magnetic moment.  The underlying difficulty is that 
Quantum Chromodynamics is strongly interacting, making it much more 
difficult to carry out computations.  At the low energies
relevant for the muon anomalous magnetic moment, the QCD coupling
becomes large and perturbation theory breaks down.  As yet,
there are no available methods for performing a first principles
computation of the hadronic contributions, though perhaps in the future
lattice gauge theory may be able to provide some useful input.  Currently,
experimental input and model calculations are used.

\begin{figure}[htb]
\begin{center}
\begin{picture}(100,38)(0,0)
\SetScale{.7}  
\SetWidth{1.4}
\ArrowLine(10,40)(45,40)
\Line(45,40.1)(100,40.1)
\ArrowLine(100,40)(130,40)
\PhotonArc(70,40)(33,0,60){4}{3}
\PhotonArc(70,40)(33,120,180){4}{3}
\GOval(70,67)(10,18)(0){.6}
\Photon(70,10)(70,40){4}{3}
\end{picture}
\end{center}
\vskip -1.5 cm 
\caption{The hadronic vacuum polarization contribution.  The wavy
lines are photons and the blob represents hadronic contributions.}
\label{HadronicVacPolFigure}
\end{figure}
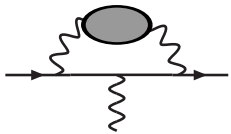

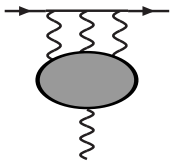
\begin{figure}[htb]
\begin{center}
\begin{picture}(100,60)(0,0)
\SetScale{.62}  
\SetWidth{1.5}
\ArrowLine(20,90)(45,90)
\Line(45,90.2)(95,90.2)
\ArrowLine(95,90)(120,90)
\Photon(50,60)(50,90){4}{2.5}
\Photon(70,62)(70,90){4}{2.5}
\Photon(90,60)(90,90){4}{2.5}
\GOval(70,48)(17,30)(0){.6}
\Photon(70,0)(70,30){4}{3}
\end{picture}
\end{center}
\vskip -1.2 cm 
\caption{The hadronic light-by-light contribution.}
\label{HadronicLbyLFigure}
\end{figure}

The two basic categories of hadronic contributions are the vacuum
polarization and light-by-light contributions depicted in
\figs{HadronicVacPolFigure}{HadronicLbyLFigure}.  The more significant
vacuum polarization contributions shown in \fig{HadronicVacPolFigure}
can be obtained from measurements of the annihilation cross section
$e^+ e^- \!\rightarrow$ hadrons, by making use of the optical theorem
and dispersion relations (see, for example, refs.~\cite{Dispersion}).
The light-by-light contributions depicted in \fig{HadronicLbyLFigure},
however, can only be obtained via model calculations with relatively
large theoretical uncertainties~\cite{HadronicLbyl}.  However, since
the latter contributions are much smaller, the vacuum polarization
uncertainty dominates.  The overall value of the light-by-light
contributions, has recently been dramatically modified because of the
discovery of a sign error in previous
calculations~\cite{HadronicLbylCorrection}; this correction helps
reduce the discrepancy between theory and experiment.

The status of the uncertainties in the hadronic corrections was
reviewed at this conference by Teubner~\cite{Teubner}.  Since the time
of the conference a number of new theoretical analyses have also
appeared~\cite{NewHadronic}. Depending on various assumptions,
currently the discrepancy is somewhere between 1.6 and 3$\sigma$.
It will, however, be some time before there is a final consensus
on its significance.

\section{CHALLENGING PERTURBATIVE COMPUTATIONS}
\label{IdeasSection}

The area of higher order perturbative computations in quantum field
theory is a rather challenging field, requiring clever ideas and
algorithms for dealing with the difficulties.  Instead of presenting a
systematic description of the methods used in state-of-the-art
computations, here we provide a selection of examples to illustrate
the type of ideas that can be helpful in challenging perturbative
calculations.  More complete descriptions may be found in various
review articles, such as refs.~\cite{QCDReviews,SmirnovBook}.

\subsection{Helicity Methods}
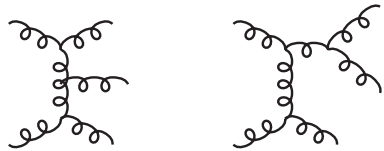
\begin{figure}[htb]
\begin{center}
\begin{picture}(190,42)(0,0)
\SetScale{.65}  
\SetWidth{1.4}
\Gluon(20,70)(50,55) {4}{2} 
\Gluon(50,55)(50,15){4}{3} 
\Gluon(50,15)(80,0){4}{2}
\Gluon(50,15)(20,0){4}{2}
\Gluon(50,55)(80,70){4}{2} \Gluon(50,35)(89,35){4}{2} 
%
%
\SetOffset(85,0)
\Gluon(20,70)(50,55) {4}{2} 
\Gluon(50,55)(50,15){4}{3} 
\Gluon(50,15)(80,0){4}{2}
\Gluon(50,15)(20,0){4}{2}
\Gluon(50,55)(75,55){4}{1.7}
\Gluon(75,55)(105,80){4}{2}
\Gluon(75,55)(105,30){4}{2}
\end{picture}
\end{center}
\vskip -1 cm 
\caption{Two of the Feynman diagrams describing the five-gluon tree
amplitude.}
\label{FiveGluonFigure}
\end{figure}

\begin{figure*}[htb]
\begin{center}
\includegraphics[width=5.3 in]{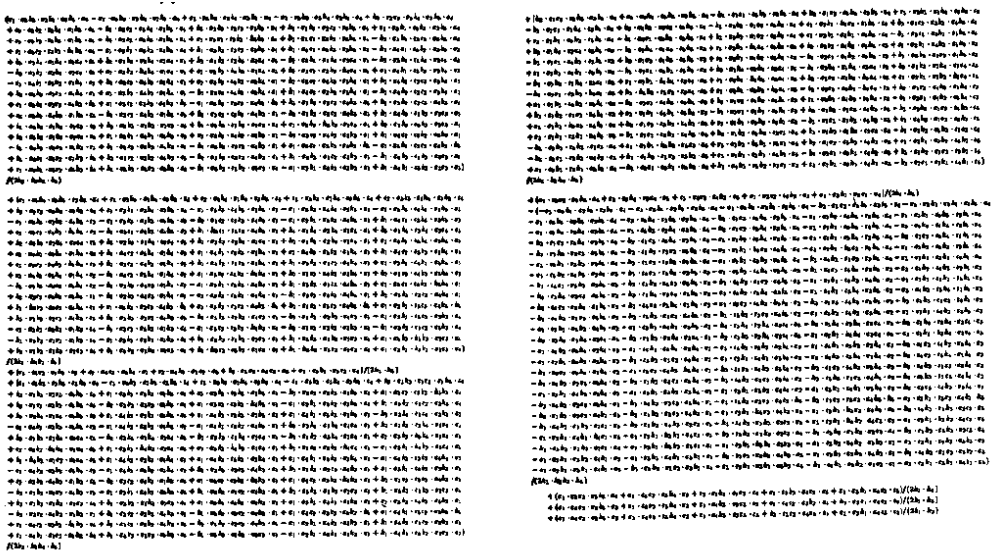}
\end{center}
\vskip -.8 cm 
\caption{The result of a brute force evaluation of the five-gluon tree
amplitude on a computer in an unreadable font. The result shown here
is expressed in terms of in terms of dot products of momentum and
polarization vectors and is the coefficient of a single color factor.}
\label{FiveGluonResultFigure}
\end{figure*}

As a first example, consider the five-gluon tree amplitude described
by Feynman diagrams of which two are depicted in
\fig{FiveGluonFigure}. These diagrams are straightforward to evaluate
by direct algebraic evaluation on a computer, using one of the
algebraic programming languages such as FORM~\cite{FORM}, MAPLE, and
so forth.  The result of evaluating the Feynman diagrams is shown in
\fig{FiveGluonResultFigure}, using a microscopic font in order to
illustrate its apparent complexity.  The result shown in this figure
is actually just the coefficient of one of the color factors; the
complete answer is much larger.

After some careful thought, a number of authors concluded that the
mess of the type shown in \fig{FiveGluonResultFigure} is not due to
any inherent property of gluon scattering but is only due to using an
inefficient representation of the amplitude~\cite{SpinorHelicity,XZC}.
A much better representation is to use helicity states instead of
polarization vectors.  For each color a gluon has only two physical
states.  A polarization vector on the other hand has four components
but with two of the degrees of freedom implicitly removed by gauge
invariance.  It is this mismatch between the number of physical states
and the number of components of the polarization vectors that is at
the origin of the apparent complexity.  A much better representation of the
amplitude may be found by making use of helicity or circular
polarization.  Using its most primitive implementation, one would
replace each polarization vector with an explicit circular
polarization:
\begin{equation}
\varepsilon_\mu^+  =  (0,1, + i, 0) \,, \hskip 1 cm 
\varepsilon_\mu^-  =  (0,1, - i, 0) \,.
\end{equation}
However, this representation is not Lorentz covariant.  Much better
Lorentz covariant representations have been
invented~\cite{SpinorHelicity}, making use of spinor representations
that are more convenient to use.  This is sometimes referred to as
`Chinese Magic' in recognition of the especially beautiful formulation
due to Xu, Zhang and Chang~\cite{XZC}.

In any case, if we make use of helicity, the complete physical content of 
\fig{FiveGluonResultFigure} can instead be written much more neatly in 
terms of the color stripped helicity amplitudes:
\begin{eqnarray}
&& \hskip -.7 cm A_5(1^\pm,2^+,3^+,4^+,5^+) = 0\,, \nonumber \\
&& \hskip -.7 cm A_5(1^-,2^-,3^+,4^+,5^+) = 
\Biggl( {s_{12}^3 \over s_{23} s_{34} s_{45} s_{51}} \Biggr)^{1/2}\! \!, 
   \label{FiveGluonTree}\\ 
&& \hskip -.7 cm A_5(1^-,2^+,3^-,4^+,5^+) = 
\Biggl( {s_{13}^3 \over s_{23} s_{34} s_{45} s_{51}} \Biggr)^{1/2}\!\!,
 \nonumber
\end{eqnarray}
where the $s_{ij} = (p_i + p_j)^2$ are the kinematic invariants appearing in
the amplitude and the $i$ and $j$ are the labels distinguishing the
different gluons.  The $+$ and $-$ superscripts for each particle $i$
label the helicity (in a convention where all legs are outgoing).

Given the remarkable cleanup of the amplitude, it is perhaps not
surprising that helicity representations have allowed seemingly
impossible computations to proceed.  Perhaps the most famous example
of this is the $n$-point generalization of eq.~(\ref{FiveGluonTree})
for special helicity configurations, obtained by Parke and
Taylor~\cite{ParkeTaylor}. They conjectured that in QCD
the leading order color-stripped $n$-gluon amplitudes satisfy,
\begin{eqnarray}
&& \hskip -.7 cm A_5(1^\pm,2^+,3^+,4^+,\ldots, n+) = 0\,, \nonumber \\
&& \hskip -.7 cm A_5(1^-,2^-,3^+,4^+,\ldots ,n^+)  \\ 
&& \hskip 2 cm 
= \Biggl( {s_{12}^3 \over s_{23} s_{34} s_{45} \ldots s_{n1}} \Biggr)^{1/2}  
\!\! ,  \label{PTequation} 
\end{eqnarray}
for the special helicity configurations indicated by the plus and
minus labels.  Using standard Feynman rules one would need to evaluate
an infinite number of diagrams to obtain this.  However, using
helicity methods together with recursive techniques, a
proof~\cite{BerendsGiele} that the Parke-Taylor amplitude are correct
was presented soon after the original conjecture. Helicity methods
have also been used for constructing infinite sequences of one-loop
amplitudes, again for cases of special helicity
configurations~\cite{Alln,Fusing}.  As a more recent example, the
first computation of a two-loop $2\rightarrow 2$ scattering amplitude
in QCD~\cite{AllPlus2} also relied on helicity methods.

\subsection{Applications of Supersymmetry}

Supersymmetry has a long history of use in constructing models of
possible extensions to the Standard Model.  Another application of
supersymmetry is providing strong checks on non-trivial calculations
in Standard Model physics.  The high degree of symmetry renders the
dynamics of supersymmetric theories far simpler than those of
non-supersymmetric ones used in describing the Standard Model.
Calculations in supersymmetric theories can therefore serve as `toys' for
devising methods for dealing with more complicated Standard Model
calculations.  Moreover a QCD calculation is a close relative of a
similar calculation in a supersymmetric version of QCD where the
quarks are replaced by gluinos.  This allows one to perform checks of
QCD calculations by performing minor alternations to the calculations
so that they become supersymmetric.

A particularly striking example of the use of supersymmetry as a check
is from the computation of the four-loop QCD $\beta$-function
performed by van Ritbergen, Vermaseren and Larin~\cite{Beta4}. The
$\beta$ function is a fundamental quantity controlling the running of
the coupling constant.  This quantity was already known earlier in a
supersymmetric version of QCD ~\cite{Jones}.  However, due to certain
technical complications having to do with the different ways that the
theories were rendered finite, the two computations could not be
compared directly. Nevertheless a non-trivial cancellation of a
particular color factor occurs when the QCD result is altered to be
supersymmetric, in agreement with the expectation from the
supersymmetric version.  This suggests that the four-loop calculation
of the QCD $\beta$-function is indeed correct.  As a more recent
example, supersymmetry identities have also been applied as checks on
two-loop calculations of scattering
amplitudes~\cite{BDDgggg,TwoloopSusy}.

\subsection{Quantum Gravity Example of Growth of Difficulty}

It is not difficult to find problems in quantum field theory which
cannot possibly be solved by brute force alone.  In many cases
systematic algorithms for evaluating Feynman diagrams are not known,
but even where algorithms are known, problems can easily surpass the
capacity of any computer.  One example of a particularly interesting
problem which possesses this property is the ultraviolet divergences
of quantum gravity.

The often-repeated statement that quantum mechanics and General
Relativity are incompatible arises from the bad ultra-violet
divergences of (super) gravity theories.\footnote{There actually is no
incompatibility in perturbation theory, up to an energy scale of
$10^{19}$ GeV, if General Relativity is viewed as an effective field
theory~\cite{Donoghue} instead of as a fundamental theory.}  There are
strong arguments that this is in fact correct based on powercounting
of the gravity Feynman diagrams (see, {\it e.g.}, ref.~\cite{HoweStelle}).

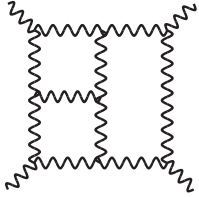
\begin{figure}
\begin{center}
\begin{picture}(100,60)(0,0)
\SetWidth{1.8}
\SetScale{.5}
\Photon(20,20)(0,0){4}{3} \Photon(140,0)(120,20){4}{3} 
\Photon(0,140)(20,120){4}{3} \Photon(120,120)(140,140){4}{3}
\Photon(20,20)(20,120){4}{9} \Photon(120,20)(20,20){4}{9}
\Photon(20,120)(120,120){4}{9} 
\Photon(120,120)(120,20){4}{9} 
\Photon(70,120)(70,20){4}{9} \Photon(20,70)(70,70){4}{5}
\end{picture}
\end{center}
\vskip -1 cm 
\caption{A sample diagram whose divergence part would need to be
evaluated in order to determine the ultra-violet divergence of a
supergravity theory. The lines represent graviton propagators and the
vertices three-graviton interactions.}
\label{GravityThreeLoopFigure}
\end{figure}

For the case of pure Einstein gravity, these arguments were
confirmed by an explicit two-loop computation performed by Sagnotti and
Goroff~\cite{Sagnotti}.  However, no proof exists for supergravity
theories.  The first potential divergence allowed by supergravity occurs at
three quantum loops. One of the Feynman diagrams which would need to
be evaluated to obtain the divergence 
is shown in \fig{GravityThreeLoopFigure}. A naive
estimate of the total number of terms in this diagram, based on
the structure of the propagators and vertices yields 
approximately $10^{21}$ terms.  If one could evaluate terms at the rate of a
billion per second, it would take approximately 20,000 years to
extract the ultra-violet divergence of just this single diagram.  It is
no surprise that this computation has not been attempted.

There are, however, methods which can bring this problem into the
realm of possible calculations.  A trick is to map the gravity
calculation in products of gauge theory calculations using certain
string theory relations between amplitudes~\cite{KLT,BDDPR}.  The net
effect of this is that an impossible problem can be mapped into an
extremely difficult problem, which also has not been solved, as yet.
This technique has, however, already been used to show that at least
for the case of maximally supersymmetric gravity the onset of
divergences is delayed until at least five quantum
loops~\cite{BDDPR,Stelle}.

\section{STATUS OF LOOP CALCULATIONS}
\label{OverviewSection}

Before surveying the main advance since the last ICHEP conference, it
is useful to survey the status of quantum loop calculations. Here we
do not discuss tree-level calculations which have also seen
considerable progress over the years.

\subsection{Status of one-loop calculations}

In 1948 Schwinger dealt with  one-loop three-point
calculations~\cite{Schwinger} such as that of the
anomalous magnetic moment of leptons described in
\sect{AnomalousMuonSection}.  It did not take very long before Karplus
and Neuman calculated light-by-light scattering in QED in their
seminal 1951 paper~\cite{KarplusNeuman}.  In 1979 Passarino and Veltman
presented the first of many systematic algorithms for dealing with
one-loop calculations with up to four external particles, leading to
an entire subfield devoted to such calculations.  Due to the
complexity of non-abelian gauge theories, however, it was not until 1986 that
the first purely QCD calculation involving four external partons was
carried out in the work of Ellis and
Sexton~\cite{EllisSexton}.  

The first one-loop five-particle scattering amplitude was then
calculated in 1993 by Lance Dixon, David Kosower and myself~\cite{FiveGluon}
for the case of five-gluon scattering in QCD.  This was
followed by calculations of the other five-point QCD
subprocesses~\cite{FiveQuark}, with the associated physical
predictions of three-jet events at hadron colliders appearing somewhat
later~\cite{Kilgore3Jet,Nagy}.  A number of other five-point
calculations have also been completed.  One example of a
state-of-the-art five-point calculation was presented in a parallel
session by Doreen Wackeroth~\cite{Wackeroth}, who described the
calculation of $p p \rightarrow \bar t t H$ at next-to-leading order
in QCD~\cite{ttH}.  This process is a useful mode for discovering the
Higgs boson as well as measurement of its properties.  Other examples
are NLO calculations for $e^+ e^- \rightarrow 4$
jets~\cite{ZJetsBDK,ZJetsGlover,ZJetsPrograms}, Higgs $\!\null + 2$
jets~\cite{HiggsJets}, and vector boson $\!\null+ 2$ jet
production~\cite{ZJetsBDK,EllisVectorBoson}, which is also important
as a background to the Tevatron Higgs search, if the jets are tagged
as coming from $b$ quarks.

Beyond five-external particles, the only calculations have been in
special cases.  By making use of advanced methods, for special
helicity configurations of the particles, infinite sequences of
one-loop amplitudes with an arbitrary number of external particles but
special helicity configurations have been obtained in a variety of
theories~\cite{Alln,Fusing}.  For the special case of maximal
supersymmetry, six-gluon scattering amplitudes have been obtained for
all helicities~\cite{Fusing}.  There has also been a recent
calculation of a six-point amplitude in the Yukawa
model~\cite{BinothAmpl}, as well as recent papers describing
properties of six-point integrals~\cite{BinothInt}.  These examples
suggest that that the technical know-how for computing general
six-point amplitudes is available, though it may be a rather
formidable task to carry it through. 
An efficient computer program for dealing with up to three jets at
hadron colliders now exists~\cite{Nagy}, suggesting that it would be
possible add one more jet, once the relevant scattering amplitudes are
calculated.  This would then give a much better theoretical handle on
multi-jet production at hadron colliders.

\subsection{Status of Higher Loop Computations}

\begin{figure}[htb]
\begin{center}
  \begin{picture}(150,25)(0,0)
  \SetScale{.8}
  \SetWidth{1.4}
  \Gluon(15,0)(50,0){3}{4}
  \CArc(75,0)(25,0,90)
  \CArc(75,0)(25,90,180)
  \CArc(75,0)(25,180,270)
  \CArc(75,0)(25,270,360)
  \Gluon(65,23)(65,-23){3}{5}
  \Gluon(75,25)(75,-25){3}{5}
  \Gluon(85,23)(85,-23){3}{5}

  \Gluon(100,0)(135,0){3}{4}
  \end{picture}
\end{center}
\vskip -.4 cm 
\caption{One of the 50,000 Feynman diagrams appearing in
the computation of the four-loop QCD $\beta$ function.}
\label{FourLoopBetaFigure}
\end{figure}
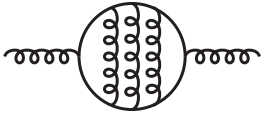

Over the years, an intensive effort has gone into calculating higher
loop Feynman diagrams.  A few samples of some impressive multi-loop
calculations are:
\begin{itemize}

\item The anomalous magnetic moment of leptons, already described in
\sect{AnomalousMuonSection}. 

\item $R = \sigma(e^+ e^- \rightarrow \hbox{hadrons})/ \sigma(e^+ e^-
\rightarrow \mu^+ \mu^-)$ which has been calculated through $
O(\alpha_s^3)$~\cite{R}. This quantity is one of the cornerstones
demonstrating that QCD is the correct theory of strong interactions.

\item The four-loop QCD $\beta$ function computed by van Ritbergen,
Vermaseren and Larin~\cite{Beta4}.  In the course of this computation 
approximately 50,000 Feynman diagrams were evaluated, of which a single one 
is displayed in \fig{FourLoopBetaFigure}.  
\end{itemize}

These types of precision calculations have been crucial for comparing
the Standard Model to experiment~\cite{PDG}. However, except for very
special cases, all two and higher-loop Feynman diagram computations
prior to a few years ago contained either one or no kinematic variables.


\section{ADVANCES OF PAST TWO YEARS}
\label{BreakthroughSection}

\begin{figure}[htb]
\begin{center}
\begin{picture}(120,25)(0,0)
\SetScale{.55}  
\SetWidth{1.8}
\Gluon(110,10)(10,10){5}{8} 
\Gluon(10,70)(110,70){5}{8}
\Gluon(50,10)(50,70){5}{4}
\ArrowLine(110,10)(110,70) \Line(110,70)(170,70) 
\Line(170,70)(170,10) \Line(110,10)(170,10)
\Photon(230,10)(170,10){5}{5} 
\Photon(230,70)(170,70){5}{5} 
\end{picture}
\end{center}
\vskip -.8 cm 
\caption{An sample process that has been evaluated. The curly lines
represent gluons, the wavy ones photons and the straight ones quarks.}
\label{TwoloopSampleFigure}
\end{figure}
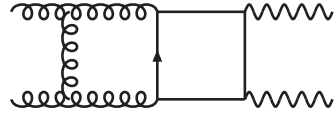

The key advance in perturbative computations over the last two years
is our ability to perform two-loop computations with more than a single
kinematic variable.  Previously, the only such calculations were for
special cases of maximal supersymmetry~\cite{BRY,BDDPR}.  The pace of
progress, due to a large extend to the influx of energetic young people
into the field, is such that even though two years ago it was not
known how to do such calculations for more general processes, now they
are commonplace.  A sample diagram illustrating the types of processes
that can now be computed is depicted in \fig{TwoloopSampleFigure}.

Every step in the construction of a physical cross-section involving a
two-loop amplitude has serious technical challenges.  Although some of
these difficulties have not been completely resolved, all aspects 
of the problem have seen remarkable progress over the last few years.

\subsection{Loop Integrals}
\label{IntegralSubsection}
\begin{figure}[htb]
\includegraphics[width=2.7 in]{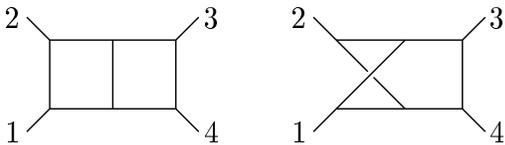}
\vskip -1 cm 
\caption{Planar and non-planar double box diagrams.}
\label{PlanarNonPlanarFigure}
\end{figure}

A crucial ingredient in this progress has been the breakthrough in
obtaining the integrals needed for computing two-loop massless $2
\rightarrow 2$ scattering
amplitudes~\cite{PBScalar,NPBScalar,IntegralsOther,IntegralReduction}.
For example, consider the scalar double box integral,
\begin{eqnarray}
&& \hskip -.8 cm \int
 {d^{D}p\over (2\pi)^{D}} \,
 {d^{D}q\over (2\pi)^{D}}\,
 {1 \over p^2\, q^2\, (p+q)^2 (p - k_1)^2 } \nonumber \\
&& \hskip -.3 cm 
\times {1 \over (p - k_1 - k_2)^2 \,
        (q - k_4)^2 \, (q - k_3 - k_4)^2 } \,.
\end{eqnarray}
which may be represented by the left diagram in
\fig{PlanarNonPlanarFigure}.  This scalar integral was a non-trivial
challenge, until it was evaluated by Smirnov~\cite{PBScalar} in terms
of standard polylogarithms.  Shortly thereafter, the non-planar scalar
integral appearing on the right-hand-side of
\fig{PlanarNonPlanarFigure} as well as other relevant integrals were
evaluated~\cite{NPBScalar,IntegralsOther}. Moreover, powerful new
methods for reducing general tensor integrals into a basis of known
integrals were developed~\cite{IntegralReduction}, making use of
integration by parts~\cite{ChetyrkinIBP} and Lorentz
invariance~\cite{LorentzIdentities}.  A number of important further
improvements have also been made allowing for more systematic
evaluations~\cite{IntegralsImproved}.  The final expressions obtained
with these methods are all in terms of standard function such as
logarithms, polylogarithms as well as generalizations of these
functions that are also amenable to efficient numerical evaluation.

\vskip -.4 cm 

\subsection{Amplitude Calculations}

The first Standard Model calculation of a $2 \rightarrow 2$ scattering
amplitude was that of gluon scattering in QCD for a special helicity
configuration~\cite{AllPlus2}.  By making use of the loop integration
breakthrough, since then many new two-loop amplitudes appeared:

\begin{itemize}
\item Bhabha scattering in QED, $e^+ e^- \rightarrow e^+ e^-$, 
in the ultra-relativistic limit~\cite{BhabhaTwoLoop}.  Bhabha scattering
is useful for monitoring luminosity at an $e^+ e^-$ collider.

\item All two-loop $2 \rightarrow 2$ QCD 
processes~\cite{GOTY2to2,BDDgggg}.  These are essential 
ingredients for constructing NNLO jet programs for hadron colliders.

\item Light-by-light scattering, $\gamma \gamma \rightarrow \gamma
\gamma$, in the ultra-relativistic limit~\cite{Lbyl}, which is of
interest for future photon-photon colliders.

\item Gluon fusion into a photon pair, $gg \rightarrow 
\gamma \gamma$~\cite{BDDgggamgam}. 
As described in
\sect{ExamplesSection}, this subprocess is important as a background to
Higgs production when a light Higgs decays into a pair of photons.

\item Two loop QED and QCD corrections to massless fermion boson
scattering~\cite{FermionBoson}: $\bar q q \rightarrow \gamma \gamma$,
$\bar q q \rightarrow g \gamma$, and $e^+ e^- \rightarrow \gamma
\gamma$.

\item $e^+ e^- \rightarrow 3$ partons~\cite{NNLOee}. This is a key
process for a future high energy $e^+ e^-$ collider and it will allow
precision measurements of strong coupling processes at the 1\% 
level~\cite{Burrows}.

\item Deeply inelastic scattering 2 jet production 
and $p p \rightarrow W,Z + 1$ jet~\cite{NNLOee,GehrmannRemiddi}.
\end{itemize}

As more integrals including ones with massive legs are worked
out~\cite{MassiveInts}, other amplitudes in QED, electroweak theory,
heavy quark physics, {\it etc.} will surely also be evaluated.  Such
calculations are just the initial steps since they are for matrix
elements and not for physical predictions.  As yet only the $\gamma
\gamma \rightarrow \gamma \gamma$ and $gg \rightarrow \gamma \gamma$
amplitudes have been implemented in phenomenological
studies~\cite{Lbyl,Hbkgd}.

\subsection{Infrared Divergences}

The source of much grief in perturbative calculations are infrared
divergences.  At the end, for any appropriately defined physical
quantity these divergences must cancel~\cite{LeeNauenberg}.  However,
at intermediate steps of calculations, before the various
contributions are combined to form a physically meaningful quantity,
severe divergences arise from soft (i.e. $p_i
\rightarrow 0$) or collinear particles.   At a given
order or perturbation theory, a physical quantity typically involves
both virtual and real emission contributions that are separately
infrared divergent.  

\begin{figure}[htb]
\begin{center}
\includegraphics[width=2.9 in]{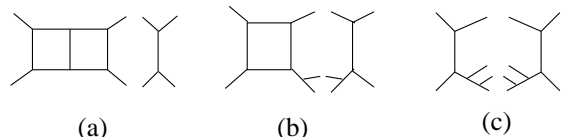}
\end{center}
\vskip -1 cm 
\caption{The various contributions at NNLO to the squared matrix elements:
 (a) virtual, (b) single real emission and (c) double real emission.}
\label{NNLOContributionsFigure}
\end{figure}

At two-loops it becomes much more difficult to deal with these
divergences due to their intricate structure.  An important result is
Catani's Magic Formula'~\cite{CataniTwoloop} which specifies the
precise form of the two-loop divergences.  The `magic' is that it was
obtained before any explicit calculations were possible.  Feynman
supposedly once remarked that before one calculates one should know
the result.  Needless to say, this ideal is difficult to achieve, but
Catani's Magic Formula is a partial implementation in the sense that
the infrared divergences that must emerge are known prior to beginning
a calculation.  Catani's Magic Formula then provides important
guidance in any explicit two loop calculation.  It also provides a
good way for organizing amplitudes into divergent parts which
ultimately drop out of physical quantities and the finite
contributions which determine the physics.

In order for the advances in computing scattering amplitudes to be
used in comparisons of theory to experiments, the various contributions need to
be combined into a numerical program for producing the plots comparing
theory to experiment.  In general, this is non-trivial because of the
infrared divergences plaguing intermediate steps of the calculations.
At NLO general methods for dealing with the infrared singularities
exist~\cite{NLOIR}, but at NNLO as yet there is no complete solution.
There is, however, a partial understanding.  In particular, the
universal behavior of amplitudes entering into NNLO calculations as
the momenta of partons become either soft or collinear has been worked
out~\cite{OneloopSplit,TreeSplit}.  Moreover, in certain special
cases, such as Drell-Yan~\cite{DrellYan}, inclusive Higgs production
at hadron colliders~\cite{HarlanderKilgore,AnastasiouMelnikov} and
$e^+ e^- \rightarrow \gamma + 1$ jet~\cite{GehrmannDeRidder} this has
been worked out.  It is now clear that more general methods for
dealing with the infrared divergences at NNLO will be constructed
in the near future.

\subsection{Parton Distribution Functions}

In order to have true NNLO predictions for observables at hadron
colliders a crucial ingredient is NNLO parton distribution functions.
At NLO the evolution kernels needed for the DGLAP
equations~\cite{DGLAP} were calculated long ago~\cite{NLOpdf}.  The
calculation of the next order has been a difficult challenge since
then, but recent years have seen some impressive progress.  The
problem is partially solved~\cite{NNLOpdfExact}, and very good
approximate solutions based on computing Mellin moments have also been
obtained~\cite{NNLOMoments}.  These are starting to be implemented in
global fits by the MRST group~\cite{MRSTNNLO}.  There is also some
related work on NNLO quark and gluon distributions inside
photons~\cite{NNLOPhoton} instead of protons.  We can anticipate that
it will not be long before all standard parton distribution functions
are fully implemented at NNLO.  (A description of other important
parton distribution function issues, such as uncertainties, may be
found in the talks of Giele and Stump~\cite{PDFUncertainty}.)

\subsection{Problems with multiple mass scales.}

For problems with multiple mass scales, in recent years there has also
been quite a bit of progress.  A powerful general strategy is known as
`the strategy of
regions'~\cite{StrategyRegionsOld,StrategyRegionsNew,SmirnovBook}.
The basic trick makes use of dimensional regularization allowing one
to series expand expressions in various kinematic limits or regions
without the need for keeping track of the boundaries between regions.
The key progress is that now there is a universal method for dealing
with this, instead of case-by-case analysis depending on boundaries
between the regions.  This has been applied to a large number of
problems~\cite{StrategyRegionsExamples}, ranging from atomic physics,
heavy quark production and decay, large electroweak logarithms and
very recently lattice gauge theory.  As one example from this
conference, the electroweak logarithms described by K\"uhn~\cite{Kuhn}
were obtained using the strategy of regions.

\subsection{Numerical Methods}

Another area where there has been recent developments is in purely
numerical techniques for evaluating Feynman diagrams.  An important
recent advance in numerical methods was made by Ghinculov and
Yao~\cite{GhinculovYao}, who produced a method for dealing with
two-loop problems where the infrared divergences are mild, as typical
in problems in electroweak theory.  An extension of this was described
in the parallel session talk of Passarino~\cite{PassarinoTwoloop},
targeting state-of-the-art electroweak calculations.  In QCD where
there are severe infrared divergence, another numerical method, by
Binoth and Heinrich~\cite{BinothNumerical}, has been applied as an
important check on state-of-the-art analytical two-loop calculations.
At this conference Dave Soper described a purely numerical method for
massless QCD amplitudes~\cite{Soper}. So far, this latter method has
been used to reproduce the seminal result for $e^+ e^- \rightarrow 3$
jets at NLO obtained by Ellis, Ross and Terrano~\cite{ERT} in 1981,
but it is not yet clear if this method will be applied to
state-of-the-art calculations.

\section{SAMPLE APPLICATIONS OF THE BREAKTHROUGH}
\label{ExamplesSection}

In the month prior to the conference, the first two applications of
the new advances in two-loop calculation to specific problems of
interest for collider physics
appeared~\cite{Hbkgd,AnastasiouMelnikov}. These first examples focus
on Higgs physics in part because of its importance for the collider
physics program, but also because the infrared singularities for these
cases are relatively straightforward to handle.  Once the general
problem of infrared divergent phase space integrals is solved, many
more applications will appear.

The first of these applications provides an improved understanding of
the QCD background to Higgs production and decay for the case of a
light Higgs.  The second application, which appeared just two weeks
prior to the conference, provides the exact result for inclusive Higgs
production~\cite{AnastasiouMelnikov} at NNLO.  This result is in
perfect agreement with an earlier series expansion due to Harlander
and Kilgore~\cite{HarlanderKilgore} and in very good agreement with an
even earlier approximate solution~\cite{CataniHiggs,GrazziniProc}.

\subsection{Inclusive Higgs Production}

In his talk, Frixione~\cite{FrixioneTalk} described the important
quest to obtain reliable predictions for Higgs production at the
Tevatron and LHC.  Previously, inclusive Higgs production had been
calculated through NLO~\cite{HiggsNLO}, leaving a substantial
theoretical uncertainty in the result.  The new NNLO calculations
significantly reduce the theoretical uncertainties.

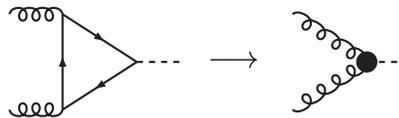
\begin{figure}[htb]
\begin{center}
\begin{picture}(140,60)(0,0)
\SetScale{.4}
\SetWidth{2.0}
\Gluon(50,30)(0,30){6}{3}
\Gluon(0,120)(50,120){6}{3}
\ArrowLine(50,30)(50,120)
\ArrowLine(120,75)(50,30)
\ArrowLine(50,120)(120,75)
\DashLine(120,75)(160,75){6}
%
\Text(75,30)[l]{\large$\longrightarrow$}
\SetOffset(87,0)
\Gluon(120,75)(50,30){6}{4}
\Gluon(50,120)(120,75){6}{4}
\DashLine(120,75)(160,75){6}
\Vertex(120,75){10}
\end{picture}
\end{center}
\vskip -1.2 cm 
\caption{For a light Higgs the top quark loop can be replaced
by an effective vertex. The curly lines represent gluons, the
solid ones the top quark, and the dashed ones the Higgs boson.}
\label{HiggsVertexFigure}
\end{figure}

Because of the large Yukawa coupling of the Higgs to top quarks and
the fact that the LHC acts as a glue factory the dominant Higgs
production mechanism LHC is though gluon fusion via a top loop.  For a
light mass Higgs boson, preferred by precision electroweak
measurements, the top loop can me replaced by a simpler effective
vertex~\cite{Vainshtein}, as depicted in \fig{HiggsVertexFigure}.  In
this case, the effective vertex is an excellent
approximation~\cite{SpiraHiggs}.

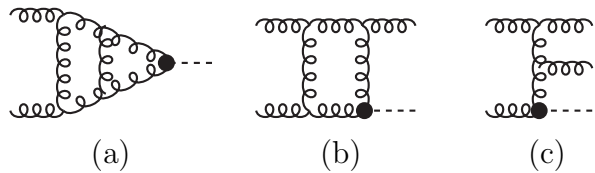
\begin{figure*}[bt]
\begin{center}
\begin{picture}(200,55)(0,0)
\SetWidth{2.0}
\SetScale{.4}
\Gluon(50,30)(0,30){6}{3}
\Gluon(0,120)(50,120){6}{3}
\Gluon(50,30)(50,120){6}{5}
\Gluon(150,75)(50,30){6}{5}
\Gluon(50,120)(150,75){6}{5}
\Gluon(90,46)(90,110){6}{3}
\DashLine(150,75)(190,75){6}
\Vertex(147,75){8}
\Text(38,-5)[c]{\small{\large (a)}}
\SetOffset(93,0)
\SetWidth{2.0}
\Gluon(50,30)(0,30){6}{3}
\Gluon(0,110)(50,110){6}{3}
\Gluon(50,30)(50,110){6}{4}

\Gluon(100,30)(50,30){6}{3}
\Gluon(50,110)(100,110){6}{3}
\Gluon(100,110)(100,30){6}{4}

\Gluon(100,110)(150,110){6}{3}
\DashLine(100,30)(150,30){6}

\Vertex(102,30){8}
\Text(32,-5)[c]{\large{(b)}}
\SetOffset(180,0)
\SetWidth{2.0}
\Gluon(50,30)(0,30){6}{3}
\Gluon(0,110)(50,110){6}{3}
\Gluon(50,30)(50,110){6}{4}
\Gluon(50,110)(100,110){6}{3}
\Gluon(50,70)(100,70){6}{3}
\DashLine(50,30)(100,30){6}
\Vertex(50,30){8}
\Text(23,-5)[c]{\small{\large (c)}}
\end{picture}
\end{center}
\vskip -.6 cm 
\caption{Sample diagrams representing (a) virtual, (b) single emission and 
(c) double emission contributions. The curly lines represent gluons, the dashed
ones Higgs bosons, and the solid circle the effective vertex coupling
gluons to the Higgs.}
\label{SampleHiggsFigure}
\end{figure*}

Using the effective Higgs vertex, the two-loop virtual contributions
of the type in \fig{SampleHiggsFigure}(a) were worked out recently by
Harlander~\cite{Harlander}.  The especially difficult part of the
calculation is dealing with the IR singular integration over the
double real emission phase space encountered when evaluating
contributions arising from diagrams of the type shown in
\fig{SampleHiggsFigure}(c).  This was solved recently by Harlander and
Kilgore~\cite{HarlanderKilgore} who constructed a series expansion of
the NNLO inclusive cross section.

The next advance was by Anastasiou and Melnikov who applied the new loop
integration technology outlined in \sect{IntegralSubsection} to obtain
the {\it exact} phase space integration over the various contributions
including the double real emission ones~\cite{AnastasiouMelnikov}.
The essential trick for doing this is based on unitarity, as encoded in
the Cutkosky cutting rules~\cite{Cutkosky}.  The relationship between
phase space and Feynman loop integrals is a rather useful trick, having, for
example, also been used to obtain a number of two-loop scattering
amplitudes~\cite{AllPlus2,BDDgggamgam,BDDgggg}.  With this trick it is
possible to replace the phase space integration, including those for
double real emission, with standard Feynman loop integrals.  Once the
problem is expressed in terms of standard Feynman integrals the new
integration methods can be applied directly.  So far this trick has been
applied to totally inclusive cross-sections, but it is very likely
that it can be extended to at least certain differential
cross-sections~\cite{AnastasiouRadcor}.

\begin{figure}[htb]
\includegraphics[width=2.7 in]{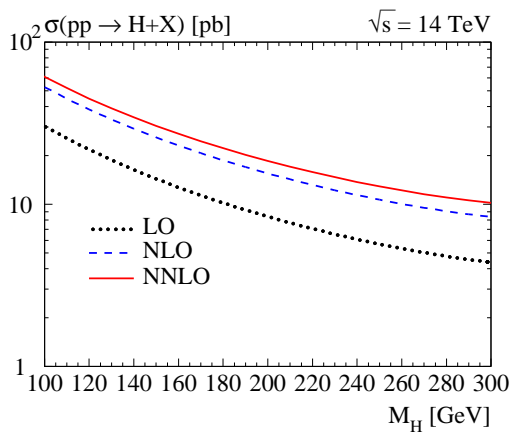}
\vskip -.7 cm 
\caption{The LO, NLO and NNLO predictions for inclusive Higgs productions at
the LHC as a functions of the Higgs mass 
(with factorization and renormalization
scales set equation to the Higgs mass).}
\label{HarlanderKilgoreFigure}
\end{figure}

The basic result of both these calculations is summarized in
\fig{HarlanderKilgoreFigure}, obtained from
ref.~\cite{HarlanderKilgore}.  The perturbative expansion clearly
stabilizes at NNLO and is reliable, given the modest shift between the
NLO and NNLO results.  Further refinements to the NNLO results should
also be forthcoming~\cite{GrazziniProc}.

\subsection{The Di-Photon Background to Higgs Decay}

The second application of the breakthrough that we describe is for the
background to observing a light ($M_H < 140$ GeV) Higgs boson at the
LHC.  For a light Higgs, a key search mode at the LHC
involves Higgs production via gluon fusion followed by the rare decay
into a pair of photons.  (For a recent survey of ways to detect 
a light Higgs boson at the LHC see, for example, ref.~\cite{MartinHiggs}.) 
Although the branching ratio is tiny, this
mode is relatively clean due to the excellent mass resolution of the
LHC detectors.  This allows the background to be measured
experimentally and subtracted from a putative
signal~\cite{HiggsExpt,Tisserand,Wielers}.  Nevertheless, it is still
important to have robust theoretical predictions in order to
systematically study the dependence of the signal relative to the
background so as to optimize Higgs search strategies.  Given that for
the case of a light Higgs boson it will take about two years of
running at the LHC before the Higgs signal is pulled out of the
background, there is good motivation for wanting to optimize search
strategies.

\begin{figure}[htb]
\begin{center}
\begin{picture}(90,45)(0,0)
\SetScale{.7} 
\SetWidth{2.0}
\Text(5,51)[l]{\Large$q$} \Text(5,0)[l]{\Large${\bar{q}}$}
\Line(20,70)(50,55) \ArrowLine(50,55)(50,15) \Line(50,15)(20,0)
\Photon(50,55)(80,70){2.5}{3} \Photon(50,15)(80,0){2.5}{3} 
\Text(59,51)[l]{\Large$\gamma$}  \Text(59,0)[l]{\Large$\gamma$}
\end{picture}
\end{center}
\vskip -.9 cm 
\caption{A leading order diagram contributing to di-photon production.}
\label{LOPhotonFigure}
\end{figure}
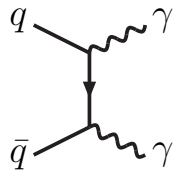

The background is composed of two pieces.  The `reducible' background
is where photons are faked by jets, or more generally by hadrons,
especially $\pi^0$s. This background can be suppressed efficiently by
photon isolation cuts, where events are rejected based on the amount
of hadronic energy near the photons.  The second source of background
is from the underlying QCD process where quarks emit photons either
directly or through fragmentation.  The process $pp \to \gamma\gamma
X$ proceeds at lowest order via the quark annihilation subprocess $q
\bar q \to \gamma\gamma$, which is independent of the strong coupling
$\alpha_s$. One of the Feynman diagrams describing this process is
shown in \fig{LOPhotonFigure}. The next-to-leading-order (NLO)
corrections to this subprocess have been incorporated into a number of
Monte Carlo programs~\cite{TwoPhotonBkgd1}, the most up-to-date being
{\tt DIPHOX}~\cite{DIPHOX}.

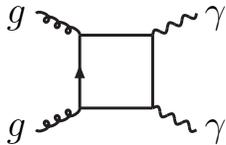
\begin{figure}[htb]
\begin{center}
\begin{picture}(70,40)(0,0)
\SetScale{.55}  
\SetWidth{2.0}
\Text(0,48)[l]{\Large$g$} \Text(0,5)[l]{\Large$g$}
\Gluon(20,90)(50,75){2.5}{3} \Gluon(50,25)(20,10){2.5}{3}
\ArrowLine(50,25)(50,75) 
\Line(50,75)(100,75) \Line(50,25)(100,25) \Line(100,75)(100,25)
\Photon(100,75)(130,90){2.5}{3} \Photon(100,25)(130,10){2.5}{3}
\Text(75,48)[l]{\Large $\gamma$}  \Text(75,5)[l]{\Large $\gamma$}
\end{picture}
\end{center}
\vskip -1 cm 
\caption{A leading order diagram contributing to the gluon fusion 
into two photons.}
\label{LOGluonFigure}
\end{figure}

Although the gluon fusion contribution depicted in \fig{LOGluonFigure}
is formally of higher order in the QCD coupling, it is enormously
enhanced by the fact that the distribution in the proton becomes very
large at small $x$.  This makes formally higher order corrections
involving gluon initial states very significant for the production of
low-mass systems ($< 200$ GeV) at the LHC.  The net result is that the
gluon fusion contribution to $pp \to \gamma\gamma X$ is comparable to the
leading-order quark annihilation
contribution~\cite{TwoPhotonBkgd1,DIPHOX,HBkgdgammagamma,ADW}.

\begin{figure}[htb]
\begin{center}
\begin{picture}(220,50)(0,0)
\SetScale{.55}  
\SetWidth{2.0}
\Text(2,48)[l]{\Large $g$} \Text(2,4)[l]{\Large $g$}
\Gluon(20,90)(50,75){2.5}{3} \Gluon(50,25)(20,10){2.5}{3}
\Gluon(50,25)(50,75){2.5}{4} \Gluon(50,75)(100,75){2.5}{4}
\Gluon(100,25)(50,25){2.5}{4}
\ArrowLine(100,25)(100,75) 
\Line(100,75)(150,75) \Line(100,25)(150,25) \Line(150,75)(150,25)
\Photon(150,75)(180,90){2.5}{3} \Photon(150,25)(180,10){2.5}{3}
\Text(101,48)[l]{\Large $\gamma$}  \Text(101,5)[l]{\Large $\gamma$}
\Text(54,-5)[c]{\large(a)}
\SetOffset(-15,0)
\Text(140,45)[l]{\Large $g$} \Text(140,8)[l]{\Large $g$}
\Gluon(270,85)(300,70){2.5}{3} \Gluon(300,30)(270,15){2.5}{3}
\Gluon(372.88,0)(338.04,17.64){2.5}{3} 
\ArrowLine(300,30)(300,70) \Line(300,70)(338.04,82.36)
\Line(338.04,82.36)(361.55,50) \Line(361.55,50)(338.04,17.64)
\Line(338.04,17.64)(300,30)
\Photon(338.04,82.36)(372.88,100){2.5}{3} 
\Photon(361.55,50)(400.60,50){2.5}{3} 
\Text(208,55)[l]{\Large $\gamma$} \Text(225,27)[l]{\Large $\gamma$}  
\Text(208,0)[l]{\Large $g$}
\Text(177,-5)[c]{\large(b)}
\end{picture}
\end{center}
\vskip -.7 cm 
\caption{Sample NLO diagrams contributing to gluon fusion 
into two photons: (a) virtual and (b) real emission contributions. }
\label{NLOGluonFigure}
\end{figure}
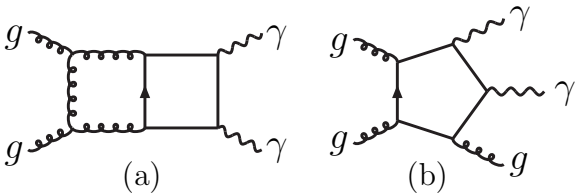

To reduce the uncertainty on the total $\gamma\gamma$ production rate,
one therefore needs to calculate the $gg \to \gamma\gamma$ subprocess
at its next-to-leading-order, even though it is formally N$^3$LO
as far as the whole process $pp \to \gamma\gamma X$ is concerned.
This involves the diagram of the type shown in 
\fig{NLOGluonFigure}.  The two-loop virtual contributions (a) were
computed recently in ref.~\cite{BDDgggamgam}, while the one-loop real
emission contribution (b) is obtained from a permutation
sum~\cite{GGGamGamG} over contributions to the one-loop five-gluon
amplitude~\cite{FiveGluon}.

To obtain the physical cross-section the two-loop contributions
of the type in \fig{NLOGluonFigure}(a) must be combined with 
the real emission contributions in \fig{NLOGluonFigure}(b) in such 
a way that all infrared singularities cancel.  In this case
the infrared singularities are all in a form which can be handled 
using standard NLO methods~\cite{NLOIR}.  Then these new contributions
must be combined with the previously obtained ones~\cite{DIPHOX}.

\begin{figure}[htb]
\begin{center}
\includegraphics[width=3. in]{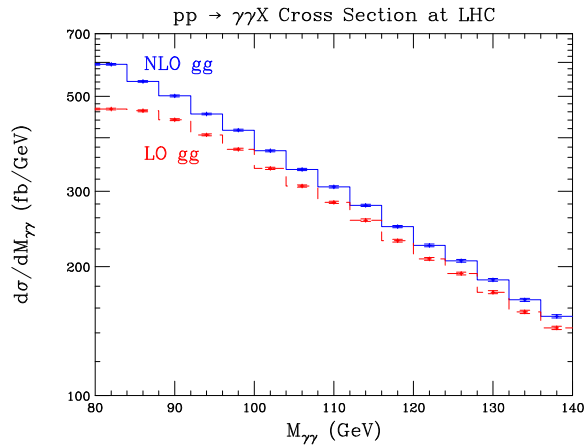}
\end{center}
\vskip -1 cm 
\caption{The $pp \rightarrow \gamma \gamma X$ cross-section as a
function of invariant mass of the di-photon pair at the LHC.  The
lower curve is obtained using {\tt DIPHOX}. The upper curve is the
result of adding in the extra NLO contributions to gluon fusion.}
\label{HiggsFigure}
\end{figure}

A sample result taken from ref.~\cite{Hbkgd} is shown in
\fig{HiggsFigure}.  The conclusion from this plot is that in the
region of interest the NLO corrections to the $gg\to\gamma\gamma$
subprocess have a relatively modest effect on the total irreducible
di-photon background to the Higgs search.  Indeed the $K$ factor
(ratio of NLO to LO contributions) for this subprocess is
significantly smaller than previous estimates somewhat enhancing the
statistical significance of the Higgs signal compared to previous
estimates.  Moreover, the relatively small shift from LO to NLO suggests
that the subprocess is under adequate theoretical
control~\cite{Hbkgd}.

A crucial question is whether we can improve the situation by finding
appropriate search strategies which enhance the signal over the background.
Some initial studies of this question at the parton level were performed
in ref.~\cite{Hbkgd}, but more realistic studies are required, including
detector effects~\cite{Wielers} and the reducible $\pi^0$ background
contributions~\cite{Tisserand,Wielers,PiBkgd}.

\section{SUMMARY AND OUTLOOK}
\label{SummarySection}

A very recent example illustrating the importance of precision
perturbative calculations is the impressive measurement of the
anomalous magnetic moment of the muon by the Brookhaven $g-2$
collaboration.  The first part of this talk outlined the intense
theoretical effort that has gone into obtaining the Standard Model
prediction at a precision matching the one of the
experiment.

This talk also highlighted the recent rapid pace of progress in our
ability to compute quantities of interest in perturbative quantum
field theory.  In particular, two-loop calculations involving up to
four external particles can now be performed in rather general cases.
This breakthrough is the culmination of years of effort.  Parallel
advances in calculations of the DGLAP evolution of parton distribution
functions have also been made.

So far the breakthrough had been applied mainly to QCD and to massless
QED, but as more types of loop integrals with masses are calculated
one can expect applications also to electroweak theory, heavy quark
physics, QED with massive particles, and so forth.  Two very recent
phenomenological applications of the advance are the calculation of
the exact next-to-next-to-leading order inclusive Higgs production
cross-section at hadron colliders and the calculation of the QCD
background to Higgs production and decay into a photon pair at the
LHC, for the case of a light Higgs.  Once general algorithms are set
up for dealing with infrared divergent phase space at
next-to-next-leading order, many more phenomenological studies will
appear.

Given the influx into the field of the many talented energetic young people 
who have contributed greatly to the advances described here, we can
be optimistic that the rapid pace of progress will continue for the
foreseeable future.

\section*{Acknowledgments}

I thank Lance Dixon and David Kosower for their many thoughtful
comments and contributions.  I also thank Bill Marciano for a number
of very illuminating discussions on the status of the anomalous
magnetic moment of the muon.  Finally, I thank Stefano Catani, Abilio
De Freitas, Einan Gardi, Bill Kilgore, Kirill Melnikov, Carl Schmidt,
and Henry Wong, as well as all the people who have made the recent
advances possible.



\begin{thebibliography}{99}

\bibitem{HiggsRadCorr}
G.~Degrassi,
arXiv:hep-ph/0102137; 
J.~Erler,
arXiv:hep-ph/0102143;
D. Abbaneo {\it et al.} [ALEPH, DELPHI, L3 and OPAL Collaborations, 
LEP Electroweak Working Group, and SLD Heavy Flavor and Electroweak Groups],
arXiv:hep-ex/0112021.

\bibitem{GM2}
H.~N.~Brown {\it et al.}  [Muon $(g-2)$ Collaboration],
Phys.\ Rev.\ Lett.\  {\bf 86}, 2227 (2001)
[arXiv:hep-ex/0102017].

\bibitem{GM22}
G.~W.~Bennett {\it et al.}  [Muon $(g-2)$ Collaboration],
Phys.\ Rev.\ Lett.\  {\bf 89}, 101804 (2002)
[Erratum-ibid.\  {\bf 89}, 129903 (2002)]
[arXiv:hep-ex/0208001].

\bibitem{Semertzidis}
Y.~K.~Semertzidis  [Muon $(g-2)$ collaboration],
in these proceedings [arXiv:hep-ph/0211038].

\bibitem{Teubner}
 T. Teubner, in these proceedings.

\bibitem{AnastasiouMelnikov}
C.~Anastasiou and K.~Melnikov,
arXiv:hep-ph/0207004.

\bibitem{HarlanderKilgore}
R.~V.~Harlander and W.~B.~Kilgore,
Phys.\ Rev.\ Lett.\  {\bf 88}, 201801 (2002)
[arXiv:hep-ph/0201206];
W.~B.~Kilgore,
in these proceedings [arXiv:hep-ph/0208143].

\bibitem{BDDgggamgam}
Z.~Bern, A.~De Freitas and L.~J.~Dixon,
JHEP {\bf 0109}, 037 (2001)
[arXiv:hep-ph/0109078].

\bibitem{Hbkgd}
Z.~Bern, L.~Dixon and C.~Schmidt,
arXiv:hep-ph/0206194, to appear in Phys.~Rev.~D.

\bibitem{GOTY2to2}
C.~Anastasiou, E.W.~Glover, C.~Oleari and M.E.~Tejeda-Yeomans,
Nucl.\ Phys.\ B {\bf 601}, 318 (2001)
[hep-ph/0010212];
%
C.~Anastasiou, E.W.~Glover, C.~Oleari and M.E.~Tejeda-Yeomans,
Nucl.\ Phys.\ B {\bf 601}, 341 (2001)
[hep-ph/0011094];
%
C.~Anastasiou, E.W.~Glover, C.~Oleari and M.E.~Tejeda-Yeomans,
Nucl.\ Phys.\ B {\bf 605}, 486 (2001)
[hep-ph/0101304];
%
E.W.~Glover, C.~Oleari and M.E.~Tejeda-Yeomans,
Nucl.\ Phys.\ B {\bf 605}, 467 (2001).

\bibitem{BDDgggg}
Z.~Bern, A.~De Freitas and L.~Dixon,
JHEP {\bf 0203}, 018 (2002)
[arXiv:hep-ph/0201161].

\bibitem{NNLOee}
L.~W.~Garland, T.~Gehrmann, E.~W.~Glover, A.~Koukoutsakis and E.~Remiddi,
Nucl.\ Phys.\ B {\bf 627}, 107 (2002)
[arXiv:hep-ph/0112081];
%
L.~W.~Garland, T.~Gehrmann, E.~W.~Glover, A.~Koukoutsakis and E.~Remiddi,
Nucl.\ Phys.\ B {\bf 642}, 227 (2002)
[arXiv:hep-ph/0206067];
%
S.~Moch, P.~Uwer and S.~Weinzierl,
arXiv:hep-ph/0207043;
%
S.~Moch, P.~Uwer and S.~Weinzierl,
arXiv:hep-ph/0210009.

\bibitem{FrixioneTalk}
S.~Frixione,
in these proceedings [arXiv:hep-ph/0211434].

\bibitem{BhabhaTwoLoop}
Z.~Bern, L.~Dixon and A.~Ghinculov,
Phys.\ Rev.\ D {\bf 63}, 053007 (2001)
[hep-ph/0010075].

\bibitem{GloverBhabha}
E.~W.~Glover, J.~B.~Tausk and J.~J.~Van der Bij,
Phys.\ Lett.\ B {\bf 516}, 33 (2001)
[arXiv:hep-ph/0106052].

\bibitem{GM2Update}
A.~Czarnecki and W.~J.~Marciano,
Nucl.\ Phys.\ Proc.\ Suppl.\  {\bf 76}, 245 (1999)
[arXiv:hep-ph/9810512];
%
A.~Czarnecki and W.~J.~Marciano,
in {\it Proc. of the 5th International Symposium on 
Radiative Corrections (RADCOR 2000) } ed. Howard E. Haber,
arXiv:hep-ph/0010194.

\bibitem{MarcianoHarbinger}
A.~Czarnecki and W.~J.~Marciano,
Phys.\ Rev.\ D {\bf 64}, 013014 (2001)
[arXiv:hep-ph/0102122].

\bibitem{Schwinger}
J.~S.~Schwinger,
Phys.\ Rev.\  {\bf 73}, 416 (1948).

\bibitem{MarcianoHadronic}
F.~Jegerlehner,
arXiv:hep-ph/0104304.
%
W.~J.~Marciano and B.~L.~Roberts,
arXiv:hep-ph/0105056.
%

\bibitem{QEDMoment}
C.~M. Sommerfield, Phys. Rev. {\bf 107},  328  (1957);
 Ann. Phys. {\bf 5},  26  (1958);
%
A. Petermann, Nucl. Phys. {\bf 3},  689  (1957);
 Helv. Phys. Acta {\bf 30},  407  (1957);
%
T.~Kinoshita,
Z.\ Phys.\ C {\bf 56}, S80 (1992);
%
V.~W.~Hughes and T.~Kinoshita,
Rev.\ Mod.\ Phys.\  {\bf 71}, S133 (1999).

\bibitem{Kinoshita4Loop}
T.~Kinoshita, B.~Nizic and Y.~Okamoto,
Phys.\ Rev.\ D {\bf 41}, 593 (1990).

\bibitem{LaportaQED}
S. Laporta and E. Remiddi, Phys. Lett. {\bf B379},  283  (1996).

\bibitem{QEDGM2}
G. Li, R. Mendel, and M.~A. Samuel, Phys. Rev. {\bf D47},  1723  (1993);
H.~H. Elend, Phys. Lett. {\bf 20},  682  (1966), erratum: ibid., vol. 21, p.
  720;
S. Laporta and E. Remiddi, Phys. Lett. {\bf B301},  440  (1993);
B. Krause, Phys. Lett. {\bf B390},  392  (1997).

\bibitem{KinoshitaCorrect}
T.~Kinoshita and M.~Nio,
arXiv:hep-ph/0210322.

\bibitem{QEDFiveLoop}
T. Kinoshita, B. Nizic, and Y. Okamoto, Phys. Rev. {\bf D41},  593  (1990);
%
A.~S. Yelkhovsky, Sov. J. Nucl. Phys. {\bf 49},  656  (1989);
%
A.~I. Milstein and A.~S. Yelkhovsky, Phys. Lett. {\bf B233},  11  (1989);
%
S.~G. Karshenboim, Phys. Atom. Nucl. {\bf 56},  857  (1993);
%
A.~L. Kataev, Phys. Lett. {\bf B284},  401  (1992);
%
A.~L. Kataev and V.~V. Starshenko, Phys. Rev. {\bf D52},  402  (1995);
%
J. Ellis, M. Karliner, M.~A. Samuel, and E. Steinfelds, hep-ph/9409376.

\bibitem{Electroweak1Loop}
K. Fujikawa, B.~W. Lee, and A.~I. Sanda, Phys. Rev. {\bf D6},  2923  (1972);
R. Jackiw and S. Weinberg, Phys. Rev. {\bf D5},  2473  (1972);
G. Altarelli, N. Cabibbo, and L. Maiani, Phys. Lett. {\bf B40},  415  (1972);
I. Bars and M. Yoshimura, Phys. Rev. {\bf D6},  374  (1972); 
W.~A. Bardeen, R. Gastmans, and B.~E. Lautrup, Nucl. Phys. {\bf B46},  315
  (1972).

\bibitem{Electroweak2Loop}
A. Czarnecki, B. Krause, and W. Marciano, Phys. Rev. D52, R2619 (1995);
A. Czarnecki, B. Krause, and W. Marciano, Phys. Rev. Lett. {\bf 76},  3267
  (1996);
T.~V. Kukhto, E.~A. Kuraev, A. Schiller, and Z.~K. Silagadze, Nucl. Phys. 
  B371,  567  (1992);
S. Peris, M. Perrottet, and E. de~Rafael, Phys. Lett. B355,  523 (1995).

\bibitem{Dispersion}
M.~Gourdin and E.~De Rafael,
Nucl.\ Phys.\ B {\bf 10} (1969) 667;
%
S.~Eidelman and F.~Jegerlehner,
Z.\ Phys.\ C {\bf 67}, 585 (1995)
[arXiv:hep-ph/9502298].

\bibitem{HadronicLbyl}
T.~Kinoshita, B.~Nizic and Y.~Okamoto,
Phys.\ Rev.\ D {\bf 31}, 2108 (1985);
%
M.~Hayakawa, T.~Kinoshita and A.~I.~Sanda,
Phys.\ Rev.\ Lett.\  {\bf 75}, 790 (1995)
[arXiv:hep-ph/9503463];
%
J.~Bijnens, E.~Pallante and J.~Prades,
Nucl.\ Phys.\ B {\bf 474}, 379 (1996)
[arXiv:hep-ph/9511388];
%
M.~Hayakawa, T.~Kinoshita and A.~I.~Sanda,
Phys.\ Rev.\ D {\bf 54}, 3137 (1996)
[arXiv:hep-ph/9601310];
%
E.~Bartos, A.~Z.~Dubnickova, S.~Dubnicka, E.~A.~Kuraev and E.~Zemlyanaya,
Nucl.\ Phys.\ B {\bf 632}, 330 (2002)
[arXiv:hep-ph/0106084].

\bibitem{HadronicLbylCorrection}
M.~Knecht and A.~Nyffeler,
Phys.\ Rev.\ D {\bf 65}, 073034 (2002)
[arXiv:hep-ph/0111058].

\bibitem{NewHadronic}
E.~de Rafael,
arXiv:hep-ph/0208251;
%
S.~I.~Eidelman, S.~G.~Karshenboim and V.~A.~Shelyuto,
arXiv:hep-ph/0209146;
%
K.~Hagiwara, A.~D.~Martin, D.~Nomura and T.~Teubner,
arXiv:hep-ph/0209187.

\bibitem{QCDReviews}
M.L.~Mangano and S.J.~Parke,
Phys.\ Rept.\ {\bf 200}, 301 (1991);
L.~Dixon,
in {\it Proceedings of Theoretical Advanced Study Institute in
Elementary Particle Physics (TASI 95)}, ed.\ D.E.\ Soper
[hep-ph/9601359];
%
Z.~Bern, L.~J.~Dixon and D.~A.~Kosower,
Ann.\ Rev.\ Nucl.\ Part.\ Sci.\  {\bf 46}, 109 (1996)
[arXiv:hep-ph/9602280];
%
M.~Steinhauser,
Phys.\ Rept.\  {\bf 364}, 247 (2002)
[arXiv:hep-ph/0201075].

\bibitem{SmirnovBook}
V.~A.~Smirnov,
{\it Applied Asymptotic Expansions In Momenta And Masses,}
{Berlin, Germany: Springer (2002) 262 p}.
%

\bibitem{FORM}
J.~A.~Vermaseren,
KEK-TH-326.

\bibitem{SpinorHelicity}
F.A.~Berends, R.~Kleiss, P.~De Causmaecker, R.~Gastmans and T.T.~Wu,
Phys.\ Lett.\ B {\bf 103}, 124 (1981);
%
P.~De Causmaecker, R.~Gastmans, W.~Troost and T.T.~Wu,
Phys.\ Lett.\ B {\bf 105}, 215 (1981).

\bibitem{XZC}
Z.~Xu, D.~Zhang and L.~Chang,
Nucl.\ Phys.\ B {\bf 291}, 392 (1987).

\bibitem{ParkeTaylor}
S.~J.~Parke and T.~R.~Taylor,
Phys.\ Rev.\ Lett.\  {\bf 56}, 2459 (1986).

\bibitem{BerendsGiele}
F.~A.~Berends and W.~T.~Giele,
Nucl.\ Phys.\ B {\bf 306}, 759 (1988);
%
D.~A.~Kosower,
Nucl.\ Phys.\ B {\bf 335}, 23 (1990).

\bibitem{Alln}
G.~Mahlon,
Phys.\ Rev.\ D {\bf 49}, 4438 (1994)
[arXiv:hep-ph/9312276];
%
Phys.\ Rev.\ D {\bf 49}, 2197 (1994)
[arXiv:hep-ph/9311213];
Z.~Bern, G.~Chalmers, L.~J.~Dixon and D.~A.~Kosower,
Phys.\ Rev.\ Lett.\  {\bf 72}, 2134 (1994)
[arXiv:hep-ph/9312333];
%
Z.~Bern, L.~J.~Dixon, D.~C.~Dunbar and D.~A.~Kosower,
Nucl.\ Phys.\ B {\bf 425}, 217 (1994)
[arXiv:hep-ph/9403226];
%
Z.~Bern, L.~J.~Dixon, M.~Perelstein and J.~S.~Rozowsky,
Nucl.\ Phys.\ B {\bf 546}, 423 (1999)
[arXiv:hep-th/9811140];
%
Z.~Bern, L.~J.~Dixon, M.~Perelstein and J.~S.~Rozowsky,
{\it Phys.\ Lett.}\ B {\bf 444}, 273 (1998)
[hep-th/9809160];
%
Z.~Bern, L.~J.~Dixon, M.~Perelstein and J.~S.~Rozowsky,
{\it Nucl.\ Phys.}\ B {\bf 546}, 423 (1999).
[hep-th/9811140].

\bibitem{Fusing}
Z.~Bern, L.~J.~Dixon, D.~C.~Dunbar and D.~A.~Kosower,
Nucl.\ Phys.\ B {\bf 435}, 59 (1995)
[arXiv:hep-ph/9409265].

\bibitem{AllPlus2}
Z.~Bern, L.~Dixon and D.A.~Kosower,
JHEP {\bf 0001}, 027 (2000)
[hep-ph/0001001].

\bibitem{Beta4}
T.~van Ritbergen, J.~A.~Vermaseren and S.~A.~Larin,
Phys.\ Lett.\ B {\bf 400}, 379 (1997).

\bibitem{Jones}
I.~Jack, D.~R.~Jones and C.~G.~North,
Nucl.\ Phys.\ B {\bf 486}, 479 (1997)
[arXiv:hep-ph/9609325].

\bibitem{TwoloopSusy}
T.~Binoth, E.~W.~Glover, P.~Marquard and J.~J.~van der Bij,
JHEP {\bf 0205}, 060 (2002)
[arXiv:hep-ph/0202266];
%
Z.~Bern, A.~De Freitas, L.~Dixon and H.~L.~Wong,
arXiv:hep-ph/0202271, to appear in Phys. Rev. D.

\bibitem{Donoghue}
J.~F.~Donoghue,
Phys.\ Rev.\ D {\bf 50}, 3874 (1994)
[arXiv:gr-qc/9405057].

\bibitem{HoweStelle}
P.~S.~Howe and K.~S.~Stelle,
Int.\ J.\ Mod.\ Phys.\ A {\bf 4}, 1871 (1989);
%
P.~S.~Howe and K.~S.~Stelle,
arXiv:hep-th/0211279.

\bibitem{Sagnotti}
M.~H.~Goroff and A.~Sagnotti,
Nucl.\ Phys.\ B {\bf 266}, 709 (1986);
%
A.~E.~van de Ven,
Nucl.\ Phys.\ B {\bf 378}, 309 (1992).

\bibitem{KLT}
H.~Kawai, D.~C.~Lewellen and S.~H.~Tye,
Nucl.\ Phys.\ B {\bf 269}, 1 (1986).

\bibitem{BDDPR}
Z.~Bern, L.~Dixon, D.~C.~Dunbar, M.~Perelstein and J.~S.~Rozowsky,
Nucl.\ Phys.\ B {\bf 530}, 401 (1998)
[hep-th/9802162].

\bibitem{Stelle}
K.~S.~Stelle,
preprint hep-th/0203015.

\bibitem{KarplusNeuman}
R.~Karplus and M.~Neuman, Phys.\ Rev.\ {\bf 83}, 776 (1951).

\bibitem{EllisSexton}
R.~K.~Ellis and J.~C.~Sexton,
Nucl.\ Phys.\ B {\bf 269}, 445 (1986).

\bibitem{FiveGluon}
Z.~Bern, L.~J.~Dixon and D.~A.~Kosower,
Phys.\ Rev.\ Lett.\  {\bf 70}, 2677 (1993)
[arXiv:hep-ph/9302280].

\bibitem{FiveQuark}
Z.~Kunszt, A.~Signer and Z.~Trocsanyi,
Phys.\ Lett.\ B {\bf 336}, 529 (1994)
[arXiv:hep-ph/9405386];
%
Z.~Bern, L.~J.~Dixon and D.~A.~Kosower,
Nucl.\ Phys.\ B {\bf 437}, 259 (1995)
[arXiv:hep-ph/9409393].

\bibitem{Kilgore3Jet}
W.~B.~Kilgore and W.~T.~Giele,
Phys.\ Rev.\ D {\bf 55}, 7183 (1997)
[arXiv:hep-ph/9610433].

\bibitem{Nagy}
Z.~Nagy,
Phys.\ Rev.\ Lett.\  {\bf 88}, 122003 (2002)
[arXiv:hep-ph/0110315].

\bibitem{Wackeroth}
S.~Dawson, L.~Orr, L.~Reina and D.~Wackeroth,
in these proceedings [arXiv:hep-ph/0210109].

\bibitem{ttH}
L.~Reina and S.~Dawson,
Phys.\ Rev.\ Lett.\  {\bf 87}, 201804 (2001)
[arXiv:hep-ph/0107101];
%
W.~Beenakker, S.~Dittmaier, M.~Kramer, B.~Plumper, M.~Spira and P.~M.~Zerwas,
Phys.\ Rev.\ Lett.\  {\bf 87}, 201805 (2001)
[arXiv:hep-ph/0107081];
%
L.~Reina, S.~Dawson and D.~Wackeroth,
Phys.\ Rev.\ D {\bf 65}, 053017 (2002)
[arXiv:hep-ph/0109066].

\bibitem{ZJetsBDK}
Z.~Bern, L.~J.~Dixon, D.~A.~Kosower and S.~Weinzierl,
Nucl.\ Phys.\ B {\bf 489}, 3 (1997)
[arXiv:hep-ph/9610370];
%
Z.~Bern, L.~J.~Dixon and D.~A.~Kosower,
Nucl.\ Phys.\ B {\bf 513}, 3 (1998)
[arXiv:hep-ph/9708239].

\bibitem{ZJetsGlover}
E.~W.~Glover and D.~J.~Miller,
Phys.\ Lett.\ B {\bf 396}, 257 (1997)
[arXiv:hep-ph/9609474];
%
J.~M.~Campbell, E.~W.~Glover and D.~J.~Miller,
Phys.\ Lett.\ B {\bf 409}, 503 (1997)
[arXiv:hep-ph/9706297].

\bibitem{ZJetsPrograms}
A.~Signer and L.~J.~Dixon,
Phys.\ Rev.\ Lett.\  {\bf 78}, 811 (1997)
[arXiv:hep-ph/9609460];
%
Phys.\ Rev.\ D {\bf 56}, 4031 (1997)
[arXiv:hep-ph/9706285];
%
Z.~Nagy and Z.~Trocsanyi,
Phys.\ Rev.\ D {\bf 59}, 014020 (1999)
[Erratum-ibid.\ D {\bf 62}, 099902 (2000)]
[arXiv:hep-ph/9806317];
%
J.~M.~Campbell, M.~A.~Cullen and E.~W.~Glover,
Eur.\ Phys.\ J.\ C {\bf 9}, 245 (1999)
[arXiv:hep-ph/9809429];
%
S.~Weinzierl and D.~A.~Kosower,
Phys.\ Rev.\ D {\bf 60}, 054028 (1999)
[arXiv:hep-ph/9901277].

\bibitem{HiggsJets}
V.~Del Duca, W.~Kilgore, C.~Oleari, C.~Schmidt and D.~Zeppenfeld,
Nucl.\ Phys.\ B {\bf 616}, 367 (2001)
[arXiv:hep-ph/0108030].

\bibitem{EllisVectorBoson}
R.~K.~Ellis and S.~Veseli,
Phys.\ Rev.\ D {\bf 60}, 011501 (1999)
[arXiv:hep-ph/9810489];
%
J.~M.~Campbell and R.~K.~Ellis,
Phys.\ Rev.\ D {\bf 62}, 114012 (2000)
[arXiv:hep-ph/0006304];
%
J.~Campbell and R.~K.~Ellis,
Phys.\ Rev.\ D {\bf 65}, 113007 (2002)
[arXiv:hep-ph/0202176].

\bibitem{BinothAmpl}
T.~Binoth, J.~P.~Guillet, G.~Heinrich and C.~Schubert,
Nucl.\ Phys.\ B {\bf 615}, 385 (2001)
[arXiv:hep-ph/0106243].

\bibitem{BinothInt}
T.~Binoth, J.~P.~Guillet and G.~Heinrich,
Nucl.\ Phys.\ B {\bf 572}, 361 (2000)
[arXiv:hep-ph/9911342];
%
T.~Binoth, G.~Heinrich and N.~Kauer,
arXiv:hep-ph/0210023;
%
A.~Ferroglia, M.~Passera, G.~Passarino and S.~Uccirati,
arXiv:hep-ph/0209219;
%
A.~T.~Suzuki, E.~S.~Santos and A.~G.~Schmidt,
arXiv:hep-ph/0210083.

\bibitem{R}
S.~G.~Gorishnii, A.~L.~Kataev and S.~A.~Larin,
Phys.\ Lett.\ B {\bf 259}, 144 (1991).
%
S.~G.~Gorishnii, A.~L.~Kataev and S.~A.~Larin,
JETP Lett.\  {\bf 53}, 127 (1991)
[Pisma Zh.\ Eksp.\ Teor.\ Fiz.\  {\bf 53}, 121 (1991)];
%
L.~R.~Surguladze and M.~A.~Samuel,
Phys.\ Rev.\ Lett.\  {\bf 66}, 560 (1991)
[Erratum-ibid.\  {\bf 66}, 2416 (1991)].

\bibitem{PDG}
K.~Hagiwara {\it et al.}  [Particle Data Group Collaboration],
Phys.\ Rev.\ D {\bf 66}, 010001 (2002).

\bibitem{BRY}
Z.~Bern, J.S.~Rozowsky and B.~Yan,
Phys.\ Lett.\ B {\bf 401}, 273 (1997)
[hep-ph/9702424].

\bibitem{PBScalar}
V.A.~Smirnov,
Phys.\ Lett.\  {\bf B460}, 397 (1999)
[hep-ph/9905323].

\bibitem{NPBScalar}
J.B.~Tausk,
Phys.\ Lett.\  {\bf B469}, 225 (1999)
[hep-ph/9909506].

\bibitem{IntegralsOther}
C.~Anastasiou, E.W.N.~Glover and C.~Oleari,
Nucl.\ Phys.\  {\bf B565}, 445 (2000)
[hep-ph/9907523];
%
C.~Anastasiou, E.W.N.~Glover and C.~Oleari,
Nucl.\ Phys.\  {\bf B575}, 416 (2000),
err. ibid.\  {\bf B585}, 763 (2000)
[hep-ph/9912251];
%
T.~Gehrmann and E.~Remiddi,
Nucl.\ Phys.\ B {\bf 601}, 248 (2001)
[arXiv:hep-ph/0008287];
%
T.~Gehrmann and E.~Remiddi,
Nucl.\ Phys.\ B {\bf 601}, 287 (2001)
[arXiv:hep-ph/0101124].

\bibitem{IntegralReduction}
V.A.~Smirnov and O.L.~Veretin,
Nucl.\ Phys.\  {\bf B566}, 469 (2000)
[hep-ph/9907385];
%
C.~Anastasiou, T.~Gehrmann, C.~Oleari, E.~Remiddi and J.B.~Tausk,
Nucl.\ Phys.\  {\bf B580}, 577 (2000)
[hep-ph/0003261].
%

\bibitem{ChetyrkinIBP}
F.~V.~Tkachov,
Phys.\ Lett.\ B {\bf 100}, 65 (1981);
%
K.~G.~Chetyrkin and F.~V.~Tkachov,
Nucl.\ Phys.\ B {\bf 192}, 159 (1981).

\bibitem{LorentzIdentities}
T.~Gehrmann and E.~Remiddi,
Nucl.\ Phys.\  {\bf B580}, 485 (2000)
[hep-ph/9912329].

\bibitem{IntegralsImproved}
S.~Laporta,
Int.\ J.\ Mod.\ Phys.\ A {\bf 15}, 5087 (2000)
[arXiv:hep-ph/0102033];
%
S.~Moch, P.~Uwer and S.~Weinzierl,
J.\ Math.\ Phys.\  {\bf 43}, 3363 (2002)
[arXiv:hep-ph/0110083].

\bibitem{Lbyl}
Z.~Bern, A.~De Freitas, L.~J.~Dixon, A.~Ghinculov and H.~L.~Wong,
JHEP {\bf 0111}, 031 (2001)
[arXiv:hep-ph/0109079].

\bibitem{FermionBoson}
C.~Anastasiou, E.~W.~Glover and M.~E.~Tejeda-Yeomans,
Nucl.\ Phys.\ B {\bf 629}, 255 (2002)
[arXiv:hep-ph/0201274].

\bibitem{Burrows}
P.~N.~Burrows {\it et al.},
arXiv:hep-ex/9612012.

\bibitem{GehrmannRemiddi}
T.~Gehrmann and E.~Remiddi,
Nucl.\ Phys.\ B {\bf 640}, 379 (2002)
[arXiv:hep-ph/0207020].

\bibitem{MassiveInts}
V.~A.~Smirnov,
Phys.\ Lett.\ B {\bf 524}, 129 (2002)
[arXiv:hep-ph/0111160].

\bibitem{LeeNauenberg}
T.~Kinoshita,
J.\ Math.\ Phys.\  {\bf 3}, 650 (1962);
%
T.~D.~Lee and M.~Nauenberg,
Phys.\ Rev.\  {\bf 133}, B1549 (1964).

\bibitem{CataniTwoloop}
S.~Catani,
Phys.\ Lett.\ B {\bf 427}, 161 (1998)
[arXiv:hep-ph/9802439].

\bibitem{NLOIR}
W.~T.~Giele and E.~W.~Glover,
Phys.\ Rev.\ D {\bf 46}, 1980 (1992);
%
W.~T.~Giele, E.~W.~Glover and D.~A.~Kosower,
Nucl.\ Phys.\ B {\bf 403}, 633 (1993)
[arXiv:hep-ph/9302225];
%
S.~Frixione, Z.~Kunszt and A.~Signer,
Nucl.\ Phys.\ B {\bf 467}, 399 (1996)
[arXiv:hep-ph/9512328];
%
S.~Catani and M.~H.~Seymour,
Nucl.\ Phys.\ B {\bf 485}, 291 (1997)
[Erratum-ibid.\ B {\bf 510}, 503 (1997)]
[arXiv:hep-ph/9605323].

\bibitem{OneloopSplit}
Z.~Bern and G.~Chalmers,
Nucl.\ Phys.\ B {\bf 447}, 465 (1995)
[arXiv:hep-ph/9503236];
%
Z.~Bern, V.~Del Duca and C.~R.~Schmidt,
Phys.\ Lett.\ B {\bf 445}, 168 (1998)
[arXiv:hep-ph/9810409];
%
D.~A.~Kosower and P.~Uwer,
Nucl.\ Phys.\ B {\bf 563}, 477 (1999)
[arXiv:hep-ph/9903515];
%
Z.~Bern, V.~Del Duca, W.~B.~Kilgore and C.~R.~Schmidt,
Phys.\ Rev.\ D {\bf 60}, 116001 (1999)
[arXiv:hep-ph/9903516];
%
S.~Catani and M.~Grazzini,
Nucl.\ Phys.\ B {\bf 591}, 435 (2000)
[arXiv:hep-ph/0007142].

\bibitem{TreeSplit}
J.M.~Campbell and E.W.N.~Glover,
Nucl.\ Phys.\ {\bf B527}, 264 (1998)
[hep-ph/9710255];
S.~Catani and M.~Grazzini,
Phys.\ Lett.\ {\bf B446}, 143 (1999)
[hep-ph/9810389];
%
S.~Catani and M.~Grazzini,
Nucl.\ Phys.\ B {\bf 570}, 287 (2000)
[arXiv:hep-ph/9908523].

\bibitem{DrellYan}
R.~Hamberg, W.~L.~van Neerven and T.~Matsuura,
Nucl.\ Phys.\ B {\bf 359}, 343 (1991).

\bibitem{GehrmannDeRidder}
A.~Gehrmann-De Ridder and E.~W.~Glover,
Nucl.\ Phys.\ B {\bf 517}, 269 (1998)
[arXiv:hep-ph/9707224].

\bibitem{DGLAP}
Yu. L. Dokshitzer, Sov. Phys. JETP {\bf 46} (1977) 641;
V.N. Gribov and L.N. Lipatov, Sov. J. Nucl. Phys. {\bf 15}
(1972) 675; 
G. Altarelli and G. Parisi, Nucl.\ Phys.\ {\bf B126} (1977) 298.

\bibitem{NLOpdf}
G.~Curci, W.~Furmanski and R.~Petronzio,
Nucl.\ Phys.\ B {\bf 175}, 27 (1980).

\bibitem{NNLOpdfExact}
E.~B.~Zijlstra and W.~L.~van Neerven,
Nucl.\ Phys.\ B {\bf 383}, 525 (1992).
%
 S. Moch and J.A.M. Vermaseren, 
                Nucl.\ Phys.\ (Proc.\ Suppl.) {\bf 89} (200) 131, 137;
S.~Moch, J.~A.~Vermaseren and M.~Zhou,
arXiv:hep-ph/0108033;
%
S.~Moch, J.~A.~Vermaseren and A.~Vogt,
preprint arXiv:hep-ph/0209100.

\bibitem{NNLOMoments}
  S.A. Larin, T. van Ritbergen, and J.A.M. Vermaseren,
                Nucl.\ Phys.\ {\bf B427} (1994) 41; 
                S.A. Larin, P. Nogueira, T. van Ritbergen, and J.A.M.
                Vermaseren, Nucl.\ Phys.\ {\bf B492} (1997) 338
S. Catani and F. Hautmann, Nucl.\ Phys.\ {\bf B427}
                (1994) 475; 
                J. Bl\"umlein and A. Vogt, Phys.\ Lett.\ {\bf B370}
                1996) 149;
                V.S. Fadin and L.N. Lipatov, Phys.\ Lett.\ {\bf B429}
                (1998) 127;
                M. Ciafaloni and G. Camici, Phys.\ Lett.\ {\bf B430}
                (1998) 349;
%
W.~L.~van Neerven and A.~Vogt,
Phys.\ Lett.\ B {\bf 490}, 111 (2000)
[arXiv:hep-ph/0007362];
%
W.~L.~van Neerven and A.~Vogt,
Nucl.\ Phys.\ B {\bf 568}, 263 (2000)
[arXiv:hep-ph/9907472];
%
W.~L.~van Neerven and A.~Vogt,
Nucl.\ Phys.\ B {\bf 588}, 345 (2000)
[arXiv:hep-ph/0006154];
%
A.~Retey and J.~A.~Vermaseren,
Nucl.\ Phys.\ B {\bf 604}, 281 (2001)
[arXiv:hep-ph/0007294].

\bibitem{MRSTNNLO}
A.~D.~Martin, R.~G.~Roberts, W.~J.~Stirling and R.~S.~Thorne,
Phys.\ Lett.\ B {\bf 531}, 216 (2002)
[arXiv:hep-ph/0201127].

\bibitem{NNLOPhoton}
S.~Moch, J.~A.~Vermaseren and A.~Vogt,
Nucl.\ Phys.\ B {\bf 621}, 413 (2002)
[arXiv:hep-ph/0110331].

\bibitem{PDFUncertainty}
W. Giele, in these proceedings; 
Daniel R. Stump, in these proceedings.

\bibitem{StrategyRegionsOld}
S.~G.~Gorishnii,
Nucl.\ Phys.\ B {\bf 319}, 633 (1989);
%
K.~G.~Chetyrkin,
Theor.\ Math.\ Phys.\  {\bf 75}, 346 (1988)
[Teor.\ Mat.\ Fiz.\  {\bf 75}, 26 (1988)];
%
V.~A.~Smirnov,
Commun.\ Math.\ Phys.\  {\bf 134}, 109 (1990).

\bibitem{StrategyRegionsNew}
F.~V.~Tkachov,
Phys.\ Lett.\ B {\bf 412}, 350 (1997)
[arXiv:hep-ph/9703424];
%
M.~Beneke and V.~A.~Smirnov,
Nucl.\ Phys.\ B {\bf 522}, 321 (1998)
[arXiv:hep-ph/9711391].

\bibitem{StrategyRegionsExamples}
K.~G.~Chetyrkin, J.~H.~Kuhn and M.~Steinhauser,
Nucl.\ Phys.\ B {\bf 482}, 213 (1996)
[arXiv:hep-ph/9606230].
%
R.~Harlander, T.~Seidensticker and M.~Steinhauser,
Phys.\ Lett.\ B {\bf 426}, 125 (1998)
[arXiv:hep-ph/9712228].
%
A.~H.~Hoang and T.~Teubner,
Phys.\ Rev.\ D {\bf 58}, 114023 (1998)
[arXiv:hep-ph/9801397].
%
K.~Melnikov and A.~Yelkhovsky,
Nucl.\ Phys.\ B {\bf 528}, 59 (1998)
[arXiv:hep-ph/9802379].
J.~H.~Kuhn, A.~A.~Penin and V.~A.~Smirnov,
Eur.\ Phys.\ J.\ C {\bf 17}, 97 (2000)
[arXiv:hep-ph/9912503].
%
M.~Beneke, A.~Signer and V.~A.~Smirnov,
Phys.\ Lett.\ B {\bf 454}, 137 (1999)
[arXiv:hep-ph/9903260].
%
I.~Blokland, A.~Czarnecki and K.~Melnikov,
Phys.\ Rev.\ D {\bf 65}, 073015 (2002)
[arXiv:hep-ph/0112267].
%
T.~Becher and K.~Melnikov,
arXiv:hep-ph/0207201.

\bibitem{Kuhn}
J.~K\"uhn, in these proceedings.

\bibitem{GhinculovYao}
A.~Ghinculov and Y.~P.~Yao,
Phys.\ Rev.\ D {\bf 63}, 054510 (2001)
[arXiv:hep-ph/0006314].

\bibitem{PassarinoTwoloop}
G.~Passarino and S.~Uccirati,
Nucl.\ Phys.\ B {\bf 629}, 97 (2002)
[arXiv:hep-ph/0112004];
%
G.~Passarino, in these proceedings.

\bibitem{BinothNumerical}
T.~Binoth and G.~Heinrich,
Nucl.\ Phys.\ B {\bf 585}, 741 (2000)
[arXiv:hep-ph/0004013].

\bibitem{Soper}
D.~E.~Soper,
Phys.\ Rev.\ D {\bf 62}, 014009 (2000)
[arXiv:hep-ph/9910292];
%
M.~Kramer and D.~E.~Soper,
Phys.\ Rev.\ D {\bf 66}, 054017 (2002)
[arXiv:hep-ph/0204113];
%
D.~E.~Soper, in these proceedings.

\bibitem{ERT}
R.~K.~Ellis, D.~A.~Ross and A.~E.~Terrano,
Nucl.\ Phys.\ B {\bf 178}, 421 (1981).

\bibitem{CataniHiggs}
S.~Catani, D.~de Florian and M.~Grazzini,
JHEP {\bf 0105}, 025 (2001)
[arXiv:hep-ph/0102227].

\bibitem{GrazziniProc}
M.~Grazzini,
in the proceedings [arXiv:hep-ph/0209302].

\bibitem{HiggsNLO}
S.~Dawson,
Nucl.\ Phys.\ B {\bf 359}, 283 (1991);
%
A.~Djouadi, M.~Spira and P.~M.~Zerwas,
Phys.\ Lett.\ B {\bf 264}, 440 (1991).

\bibitem{Vainshtein}
M.~A.~Shifman, A.~I.~Vainshtein, M.~B.~Voloshin and V.~I.~Zakharov,
Sov.\ J.\ Nucl.\ Phys.\  {\bf 30}, 711 (1979)
[Yad.\ Fiz.\  {\bf 30}, 1368 (1979)].

\bibitem{SpiraHiggs}
M.~Spira,
Fortsch.\ Phys.\  {\bf 46}, 203 (1998)
[arXiv:hep-ph/9705337].

\bibitem{Harlander}
R.~V.~Harlander,
Phys.\ Lett.\ B {\bf 492}, 74 (2000)
[arXiv:hep-ph/0007289].

\bibitem{Cutkosky}
R.E.\ Cutkosky, J.\ Math.\ Phys. {\bf 1}, :429 (1960).

\bibitem{AnastasiouRadcor}
C.~Anastasiou, L.~Dixon and K.~Melnikov,
arXiv:hep-ph/0211141.

\bibitem{MartinHiggs}
A.~De Roeck, V.~A.~Khoze, A.~D.~Martin, R.~Orava and M.~G.~Ryskin,
Eur.\ Phys.\ J.\ C {\bf 25}, 391 (2002)
[arXiv:hep-ph/0207042].

\bibitem{HiggsExpt}
CMS collaboration,
``CMS: The electromagnetic calorimeter, technical design report,''
report CERN/LHCC 97-33, CMS-TDR-4;
%
ATLAS collaboration,
``ATLAS detector and physics performance, technical design report,'' 
vol. 2, report CERN/LHCC 99-15, ATLAS-TDR-15.

\bibitem{Tisserand}
V.~Tisserand, 
``The Higgs to two photon decay in the ATLAS detector,''
talk given at the VI International Conference on Calorimetry in
High-Energy Physics, Frascati (Italy), June, 1996, LAL 96-92; Ph.D. thesis,
LAL 97-01, February, 1997.

\bibitem{Wielers}
M.~Wielers, ``Isolation of photons,'' report ATL-PHYS-2002-004.

\bibitem{TwoPhotonBkgd1}
E.L.~Berger, E.~Braaten and R.D.~Field,
Nucl.\ Phys.\ B {\bf 239}, 52 (1984); 
P.~Aurenche, A.~Douiri, R.~Baier, M.~Fontannaz and D.~Schiff,
Z.\ Phys.\ C {\bf 29}, 459 (1985); 
B.~Bailey, J.F.~Owens and J.~Ohnemus,
Phys.\ Rev.\ D {\bf 46}, 2018 (1992); 
B.~Bailey and J.F.~Owens,
Phys.\ Rev.\ D {\bf 47}, 2735 (1993); 
B.~Bailey and D.~Graudenz,
Phys.\ Rev.\ D {\bf 49}, 1486 (1994)
[arXiv:hep-ph/9307368]; 
C.~Balazs, E.L.~Berger, S.~Mrenna and C.-P.~Yuan,
Phys.\ Rev.\ D {\bf 57}, 6934 (1998)
[arXiv:hep-ph/9712471]; 
C.~Balazs and C.-P.~Yuan,
Phys.\ Rev.\ D {\bf 59}, 114007 (1999)
[Erratum-ibid.\ D {\bf 63}, 059902 (1999)]
[arXiv:hep-ph/9810319];
T.~Binoth, J.P.~Guillet, E.~Pilon and M.~Werlen,
Phys.\ Rev.\ D {\bf 63}, 114016 (2001)
[arXiv:hep-ph/0012191]; 
T.~Binoth,
arXiv:hep-ph/0005194.

\bibitem{DIPHOX}
T.~Binoth, J.P.~Guillet, E.~Pilon and M.~Werlen,
Eur.\ Phys.\ J.\ C {\bf 16}, 311 (2000)
[arXiv:hep-ph/9911340].

\bibitem{HBkgdgammagamma}
R.K.~Ellis, I.~Hinchliffe, M.~Soldate and J.J.~van der Bij,
Nucl.\ Phys.\ B {\bf 297}, 221 (1988).

\bibitem{ADW}
L.~Ametller, E.~Gava, N.~Paver and D.~Treleani,
Phys.\ Rev.\ D {\bf 32}, 1699 (1985); 
%
D.A.~Dicus and S.S.D.~Willenbrock,
Phys.\ Rev.\ D {\bf 37}, 1801 (1988).

\bibitem{GGGamGamG}
D.~de~Florian and Z.~Kunszt,
Phys.\ Lett.\ B {\bf 460}, 184 (1999)
[arXiv:hep-ph/9905283];
%
C.~Balazs, P.~Nadolsky, C.~Schmidt and C.-P.~Yuan,
Phys.\ Lett.\ B {\bf 489}, 157 (2000)
[arXiv:hep-ph/9905551].

\bibitem{PiBkgd}
T.~Binoth, J.P.~Guillet, E.~Pilon and M.~Werlen,
arXiv:hep-ph/0203064.

\end{thebibliography}
\end{document}